\documentclass[mca,article,submit,pdftex,moreauthors]{MymdpiNoLineNumbers} 

\usepackage{graphicx}
\definecolor{Red}{rgb}{0.9,0.2,0.1}
\definecolor{dkRed}{rgb}{0.5,0.2,0.4}
\definecolor{Green}{rgb}{0.2,0.9,0.2}
\definecolor{dkGreen}{rgb}{0.1,0.8,0.1}
\definecolor{Yellow}{rgb}{1,1,0}
\definecolor{Navy}{rgb}{0.1,0.1,0.4}
\definecolor{Navy2}{rgb}{0.1,0.1,0.5}
\definecolor{Black}{rgb}{0,0,0}
\definecolor{Orange}{rgb}{1,0.55,0.02}
\definecolor{Pink}{rgb}{0.88,0.09,0.77}

\definecolor{Grey}{rgb}{0.7,0.7,0.7}
\definecolor{Math}{rgb}{0.07,0.63,0.08}
\definecolor{Violet}{rgb}{0.4,0,0.7}

\firstpage{1} 
\pubvolume{1}
\issuenum{1}
\articlenumber{0}
\pubyear{2023}
\copyrightyear{2023}
\externaleditor{Academic Editor(s): Mehmet Pakdemirli}
\datereceived{26 July 2024} 
\daterevised{26 September 2024} % Comment out if no revised date
\dateaccepted{27 September 2024  } 
\datepublished{ } 
\hreflink{https://doi.org/} % If needed use \linebreak
\Title{Using Symmetries to Investigate the Complete Integrability, 
Solitary Wave Solutions and Solitons of the Gardner Equation}

\TitleCitation{Using Symmetries to Investigate the Complete  
Integrability, Solitary Wave Solutions and Solitons of the Gardner Equation}

\Author{Willy Hereman $^{1,\ddagger}$*\orcidA{} and 
\"{U}nal G\"{o}kta\c{s}  $^{2,\ddagger}$}

\AuthorNames{Willy Hereman and \"{U}nal G\"{o}kta\c{s}}

\AuthorCitation{Hereman, W.; G\"{o}kta\c{s}, \"{U}.}
\address{%
$^{1}$ \quad Department of Applied Mathematics and Statistics,
Colorado School of Mines,
Golden CO 80401-1887, USA.; whereman@mines.edu\\
$^{2}$ \quad Department of Computer Science and Engineering,
Texas A\&M University,
College Station, TX 77843-3112, USA.; ugoktas@tamu.edu}

\corres{Correspondence: whereman@mines.edu; Tel.: 
+1 303-273-3881 (W.H.)}

\secondnote{These authors contributed equally to this work.}
\abstract{
In this paper, using a scaling symmetry, it is shown how to compute 
polynomial conservation laws, generalized symmetries, recursion operators, 
Lax pairs, and bilinear forms of polynomial nonlinear partial differential 
equations, thereby establishing their complete integrability. 
The Gardner equation is chosen as the key example, as 
it comprises both the Korteweg--de Vries and modified Korteweg--de Vries equations.
The Gardner and Miura transformations, which connect these equations, are
also computed using the concept of scaling homogeneity. 
Exact solitary wave solutions and solitons of the Gardner equation
are derived using Hirota's method and other direct methods.  
The nature of these solutions depends on the sign of 
the cubic term in the Gardner equation and the underlying 
mKdV equation.  
It is shown that flat (table-top) waves of large amplitude only  
occur when the sign of the cubic nonlinearity is negative (defocusing case), 
whereas the focusing Gardner equation has  standard elastically colliding solitons.
This paper's aim is to provide a review of the integrability properties and 
solutions of the Gardner equation and to illustrate the 
applicability of the scaling symmetry approach.   
The methods and algorithms used in this paper have been 
implemented in {\em Mathematica}, but can be adapted for major computer 
algebra systems. 
}

\begin{document}

%%%%%%%%%%%%%%%%%%%%%%%%%%%%%%%%%%%%%%%%%%
% \setcounter{section}{-1} %% Remove this when starting to work on the template.

%%%%%%%%%%%%%%%%%%%%%%%%%%%%%%%%%%%%%%%%%%
% INTRO 
\section{Introduction}
\label{introduction}
Several physically important nonlinear partial differential equations (PDEs) are 
completely integrable by the inverse scattering transform (IST) method, which can be viewed as the nonlinear analog of the Fourier transform method
(see
% MDPI: According to our rules, refs should be arrange in number order. 
% we revised them in order. please confirm.
% WH: Good. We will check that the sequence of the numbers did not get 
% messed up.
~\cite{ablowitz-clarkson-book-1991,ablowitz-segur-book-1981,remoissenet-book-1999} 
and references therein).
Arguably, the~most well-known completely integrable PDEs are the Korteweg--de Vries (KdV), modified KdV (mKdV), nonlinear Schr\"{o}dinger, and~sine-Gordon equations.  
They model wave phenomena in a wide range of applications in modern theoretical and mathematical physics, including fluid dynamics, nonlinear optics, and~plasma physics, 
to name a~few.
 
``Complete integrability'' is an elusive term~\cite{zhakharov-book-1991}, but completely
integrable equations have remarkable properties and a rich mathematical structure. 
For instance, they possess infinitely many conservation laws and high-order symmetries.
They admit a Lax pair where the nonlinear PDE is replaced by a pair of linear equations whose compatibility only holds on solutions of the nonlinear PDE. 
They can be viewed as infinite dimensional Hamiltonian systems with two or three 
different Hamiltonians. 
By applying a change of dependent variable, completely integrable PDEs can be 
transformed into equations that are homogeneous of degree (second or higher) in that 
new variable and be recast in bilinear form in terms of Hirota operators~\cite{hereman-goktas-springer-2024,hirota-book-2004}. 
Most importantly, completely integrable PDEs have {\em solitary waves solutions} 
that maintain their shape and speed while propagating at a constant velocity, 
and {\em soliton solutions} made up of such solitary waves that collide elastically. 
Nowadays, the~terms solitary wave and soliton are used interchangeably, 
without reference to the elastic scattering~property.    

To obtain an initial idea about the possible complete integrability of a PDE, 
one should check if the equation has the Painlev\'{e} property~\cite{ablowitz-clarkson-book-1991,ablowitz-segur-book-1981},  
meaning that its solutions have no worse singularities than movable poles. 
To do so, one runs the so-called Painlev\'{e} test~\cite{baldwin-hereman-jnlmp-2006},
which is algorithmic and can be performed with our {\em Mathematica}
% MDPI: Please confirm if the italics should be retained. same as below.
% WH: retain italics 
code \verb|PainleveTest.m|~\cite{baldwin-hereman-code-painlevetest-2001}.  
If a PDE passes the Painlev\'{e} test, there is no guarantee that it is completely integrable,
but it is more likely than not to have special properties, viz., 
conservation laws, generalized symmetries, and~so~on. 

In this paper we use the concept of scaling homogeneity 
to symbolically compute conserved densities, fluxes, higher-order symmetries, 
Lax pairs, bilinear forms, and~Miura-type transformations for polynomial systems 
of PDEs, thereby testing their complete integrability in a variety of ways.  
To keep the paper accessible to non-experts, we only cover PDEs in $(1+1)$ dimensions, i.e.,~one space variable $(x)$ in addition to time $(t)$. 
To illustrate the steps of the computations we focus on the Gardner equation, which 
comprises the KdV and mKdV equations as special cases. 
Therefore, the~results for these two important soliton equations are obtained 
without extra effort. 
This paper's purpose is to provide a review of the integrability properties 
and solutions of the Gardner equation and illustrate the applicability 
of the scaling symmetry approach for investigating the complete integrability 
of polynomial nonlinear~PDEs.  

Scaling homogeneity is a feature common to many nonlinear PDEs,
and it provides an elegant 
way to find, e.g.,~densities and higher-order symmetries.
Indeed, these scalar quantities can be derived as linear combinations with undetermined (constant) coefficients of scaling homogeneous polynomial terms in the 
dependent variables and their derivatives.
Since the defining equations for these quantities are {\em linear}, 
the method comes down to solving linear systems for the 
undetermined~coefficients. 

For conservation laws, the~time derivative of a candidate density must 
be the total derivative of an unknown flux. 
To test this type of ``exactness'' we use the Euler operator from the calculus of variations.  
An automated computation of the corresponding fluxes requires integration by parts 
on the so-called jet space. 
In essence, one needs to integrate expressions involving {\em arbitrary} functions. 
Unfortunately, computer algebra systems (CAS) often fail at that task.  
To circumvent this shortcoming, we use the homotopy operator from 
differential geometry, which reduces the problem to a one-dimensional integral 
with respect to an auxiliary~parameter.

The existence of a sufficiently large number of non-trivial densities suffices to establish complete integrability of a given PDE. 
In most cases, the~corresponding fluxes are not needed. 
However, for~some equations, e.g.,~the Kadomtsev--Petviashvili 
equation~\cite{ablowitz-clarkson-book-1991,ablowitz-segur-book-1981}, 
it is necessary to swap the independent variables to be able to rewrite the PDE as a system of evolution equations. 
In doing so, the~roles of density and flux become interchanged, and that is why we also show the computation of the~flux. 

The computation of higher-order symmetries is performed with the same methodology without having to rely on the relationship between symmetries and conservation laws, as~expressed through Noether's theorem.
The scaling symmetry argument 
and the method of undetermined coefficients 
also apply to recursion operators, which connect the generalized symmetries.
Of course, the~recursion operator is hard to compute because it involves total 
derivatives and anti-derivatives.
Fortunately, the~defining equation for the recursion operator is linear, which again 
reduces the problem to solving a linear system.   
By applying the recursion operator to a low-order generalized symmetry, one can 
generate all higher-order symmetries one after another. 
Thus, the existence of a recursion operator provides proof that a nonlinear 
PDE is completely integrable.
With the recursion operator, one can build a hierarchy of completely 
integrable equations which have the same properties as the original 
nonlinear PDE.
For example, the~well-known Lax equation~\cite{goktas-hereman-jsc-1997,hereman-goktas-springer-2024},
which is of fifth order in $x$, is the first higher-order generalized symmetry 
of the KdV equation~\cite{goktas-hereman-acm-1999}. 
When encountering a allegedly ``new'' nonlinear PDE of high-order in the 
literature, one should verify that it does not belong to the hierarchy of 
symmetries of a well-studied PDE of lower~order.  

The computation of Lax pairs based on scaling homogeneity is more challenging
because the defining equation is no longer linear, and thus the 
method of undetermined coefficients leads to 
an algebraic system with quadratic terms. 
It is still quite straightforward to solve; however, if~parameters are present, finding 
the conditions of the parameters for which solutions exist is substantially harder.   
The knowledge of a Lax pair is essential to apply the IST, which allows one to solve 
the initial value problem for a nonlinear PDE and, consequently, find solitary wave solutions and solitons.  
Obtaining a Lax pair is the first step in the application of the IST and Riemann--Hilbert method to find soliton solutions. 
Using the Lax pair one can also derive conservation laws and~many other properties 
of the nonlinear~PDE. 

The same thing happens when applying the scaling homogeneity method to compute 
the Miura and Gardner transformations.  
Here, one has to solve quadratic systems for the undetermined coefficients. 
The Miura transformation, which connects the KdV and mKdV equations, had a profound 
impact on the discovery of conservation laws of both equations and, in~general, 
the early development of the notion of complete 
integrability of PDEs and soliton~theory. 

To apply Hirota's direct method for finding solitary waves and solitons, the~given 
PDE must be cast in bilinear form.  
Finding that bilinear representation is a non-trivial task which requires some 
insight, experience, and~often ingenuity as explained in~\cite{hereman-goktas-springer-2024}, 
our scaling-homogeneity approach is still helpful, but has its limitations---
in particular, if~finding a bilinear representation requires splitting 
an expression into two parts in such a way that each part can be written 
in bilinear form. 
This is exactly what must be conducted with regard to the mKdV and Gardner equations and, 
as far as we know, there is no algorithmic way to achieve this. 
Once the bilinear representation is found, special solutions, including solitary waves 
and solitons, can be computed~algorithmically. 

As we will also show in Section~\ref{bilinear-form}, the~Gardner equation can be 
% WH 08/03/2024 Galilean is perhaps the better name 
transformed into the mKdV equation via a Galilean transformation.
Consequently, any time new solutions of the mKdV equation are discovered, 
one obtains additional solutions to the Gardner  
equation, and~vice~versa~\cite{kirane-etal-csf-2024}.
The search for new solutions, in~particular for~the defocusing mKdV equation, 
is still an active area of research 
(see, e.g.,~refs.
% MDPI: We added ref. for the citation at the beginning. please confirm. 
% same as below.
% WH: Good. 
~\cite{mucalica-pelinovsky-lmp-2024,slyunyaev-jetp-2001,slyunyaev-pelinovsky-prl-2016,zhang-yan-physicad-2020}).
Over the last two decades, our research team has designed and implemented 
fast algorithms~\cite{baldwin-hereman-code-recoper-2003,goktas-hereman-code-invarsym-1997,poole-hereman-code-conslawsmd-2009} 
to test the integrability of nonlinear PDEs based on the 
concept of scaling \mbox{symmetry~\cite{goktas-hereman-jsc-1997,goktas-hereman-acm-1999,hereman-goktas-wester-1999,hereman-etal-cpc-1998,hereman-etal-book-birkhauser-2005,hereman-etal-book-nova-2009,poole-hereman-jsc-2011}. }
In addition, we have implemented a simplified version of Hirota's method~\cite{goktas-hereman-code-solitons-2023,hereman-goktas-springer-2024} 
and other direct methods~\cite{baldwin-hereman-code-exactsols-2003,baldwin-etal-jsc-2004}
to find exact solutions for nonlinear PDEs. 
As a matter of fact, the~computations in this paper have been largely performed with 
our software packages, which are written in {\em Mathematica} syntax, but could be adapted for other computer algebra systems such 
as {\em Maple} and {\em REDUCE}.

The paper is organized as follows. 
In Section~\ref{gardner-equation} we introduce the Gardner equation and 
mention some of its applications. 
As shown in Section~\ref{painleve-analysis}, the~Gardner equation passes 
the Painlev\'{e} test, which indicates that the equation likely 
has many interesting properties.  
In Section~\ref{scaling-symmetry}, we discuss the scaling symmetry of the 
Gardner equation, as well as the KdV and mKdV equations. 
Section~\ref{conservation-laws} is devoted to the computation of conservation laws. 
\mbox{Sections~\ref{generalized-symmetries} and~\ref{recursion-operator}} cover the computation
of generalized symmetries and the recursion operator that connects them. 
In Section~\ref{lax-pair}, we turn our attention to the computation of a Lax pair 
for the Gardner equation. 
The derivation of a bilinear representation is covered 
in Section~\ref{bilinear-form}. 
The computation of the Gardner transformation (connecting solutions 
of the Gardner equation and KdV equations) is shown in Section~\ref{gardner-transformation}.

With the important quantities of the Gardner equation being computed, we show in Sections~\ref{kdv-equation} and~\ref{mkdv-equation} how our results translate 
to the KdV and mKdV equations.
Solitary wave solutions and solitons are covered in Sections~\ref{gardner-solitary-wave-solutions} and~\ref{gardner-solitons}, 
where we show that the nature of the solutions of the Gardner and mKdV equations 
depends on the sign of the cubic nonlinearity.  
A brief discussion of our software packages used in this study is given in Section~\ref{symbolic-software}. 
Finally, in~Section~\ref{conclusions-future-work}, we draw some conclusions 
and mention a few topics for future~work. 

% GARDNER EQUATION 
\section{The Gardner~Equation}
\label{gardner-equation}
The Gardner equation~\cite{gardner-etal-commpureapplmath-1974} is the 
{\em nonlinear} PDE 
% MDPI: The format of each character (italic or not) in an Equation should be 
% consist with the main text, please check and revise in all tex 
% if necessary
% WH: Good. 
\begin{equation}
\label{gardner} 
u_t + \alpha u u_x + \beta u^2 u_x + u_{3x} = 0,
\end{equation}
where $u(x,t)$ is the dependent variable (or field) 
which is a function of space variable $x$ and time $t$.
The subscripts denote partial derivatives, 
i.e., $u_t = \frac{\partial u}{\partial t}, u_x = \frac{\partial u}{\partial x},$
and $ u_{3x} = \frac{\partial^3 u}{\partial x^3}.$ 
The parameters $\alpha$ and $\beta$ are real numbers for which  
the values $\pm 1$ and $\pm 6$ are frequently used in the~literature.

The Gardner equation has the KdV and mKdV equations as special cases. 
Therefore, (\ref{gardner}) is sometimes called the combined or mixed 
KdV-mKdV equation. 
Indeed, for~$\beta = 0$, (\ref{gardner}) reduces to the ubiquitous 
Korteweg--de Vries equation~\cite{korteweg-devries-philmag-1895},
\begin{equation}
\label{kdv} 
u_t + \alpha u u_x + u_{3x} = 0,
\end{equation}
which models, e.g.,~shallow water waves~\cite{hereman-encyclopia-2008}, 
ion-acoustic waves in plasmas~\cite{ablowitz-clarkson-book-1991,ablowitz-segur-book-1981,drazin-johnson-book-1989},
and many other nonlinear wave phenomena where solitons arise. 
When $\alpha = 0$, (\ref{gardner}) becomes the mKdV equation~\cite{ablowitz-clarkson-book-1991,gardner-etal-commpureapplmath-1974},
\begin{equation}
\label{mkdv} 
u_t + \beta u^2 u_x + u_{3x} = 0,
\end{equation}
which models internal ocean waves, electromagnetic surface waves, waves in plasmas, 
and more~\cite{ablowitz-clarkson-book-1991}.  

% WH Oct 2, 2024 correct and order also correct 
Equation~(\ref{gardner}) appears for the first time in~\cite{miura-i-jmp-1968,miura-etal-ii-jmp-1968} in the context of the Miura 
transformation (see Section~\ref{gardner-transformation}), 
which connects the mKdV and KdV equations.  
The Gardner equation has a long history~\cite{gardner-etal-commpureapplmath-1974,kakutani-yamasaki-jpsjpn-1978} 
and many applications~\cite{zhang-etal-revmathphys-2014} in fluid dynamics:
in particular, for~modeling long internal water waves~\cite{grimshaw-studapplmath-2015,miles-tellus-1979}, 
the dynamics of undular bores~\cite{kamchatnov-etal-pre-2012}, 
and waves in multi-species plasma physics~\cite{olivier-etal-physplasmas-2020,olivier-etal-jplasmaphys-2016,torven-physrevlett-1981,verheest-etal-jplasmaphys-2024}.
There is a wealth of information available about the Gardner equation. 
The equation appears in most books about solitons and integrability 
(see, e.g.,~ref.~\cite{hereman-encyclopia-2008}, which includes a list of soliton books).

Without loss of generality, we take $\alpha \ge 0$ in (\ref{gardner}), because 
if $\alpha < 0$, replacing $u$ by $-u$ would make the coefficient of $u u_x$ 
positive again. 
However, no discrete symmetries of $x, t,$ or $u$ will flip the sign of the 
coefficient of $u^2 u_x$, so the cases for positive and negative $\beta$ will have to 
be treated separately.  
Exact solutions of (\ref{gardner}) with $\beta > 0$, called the 
{\em focusing} Gardner equation, are quite different from these of the 
{\em defocusing} case, where $\beta < 0$.

We focus on (\ref{gardner}) because it is a typical example of a scalar 
$(1+1)$-dimensional \mbox{evolution equation,}
\begin{equation}
\label{scalarpde} 
u_t = F(x,t,u,u_x,u_{xx}, \cdots, u_{nx}),
\end{equation}
of order $n$ in $x$ and first order in $t$, with~$n = 3$ in (\ref{gardner}). 
More importantly, many of the results for (\ref{gardner}) lead to the 
corresponding results for the KdV and mKdV equations by setting 
$\beta = 0$ or $\alpha = 0$, respectively. 
Like the latter two equations, the~Gardner equation is known to be completely 
integrable for both signs of $\beta$, but the solutions are quite different 
for $\beta > 0$ and $\beta < 0$. 
The latter case is the hardest to deal with. 
The solitary wave solutions and solitons for the focusing Gardner equation 
follow from those of a focusing mKdV equation to which the Gardner equation 
can be reduced. 
The so-called ``table-top'' or ``flat-top'' soliton, corresponding 
to a large amplitude, only exists for the defocusing 
Gardner~equation.  

In some applications, the coefficients $\alpha$ and $\beta$ are time dependent
~\cite{grimshaw-studapplmath-2015}. 
In that case, (\ref{gardner}) has only a couple of conservation laws 
(conservation of mass and wave action flux, see Section~\ref{conservation-laws}) 
and is non-integrable.
The treatment of (\ref{gardner}) with varying coefficients is outside the 
scope of this~paper. 

% PAINLEVE ANALYSIS 
\section{Painlev\'{e} Analysis}
\label{painleve-analysis}
In this Section we check if (\ref{gardner}) has the Painlev\'{e} property. 
Performing the Painlev\'{e} test is straightforward but usually involves 
lengthy computations best performed with a software package such as 
\verb|PainleveTest.m| 
\cite{baldwin-hereman-code-painlevetest-2001}.
In essence, one investigates a Laurent series solution for (\ref{gardner}),
\begin{equation}
\label{gardnerlaurentseries}
u(x,t) = g(x,t)^{\sigma} \sum_{k=0}^{\infty} u_k(x,t) g(x,t)^k, 
\end{equation}
in which the coefficients $u_k(x,t)$ are analytic functions in a neighborhood 
of the singular manifold $g(x,t)$. 
Furthermore, $g(x,t) = 0$ determines the poles and $g(x,t)$ is assumed to be non-characteristic (i.e., $g_x(x,t) \ne 0$).  
The negative integer $\sigma$ determines the leading order term, $u_0 \, g^{\sigma}$,
in (\ref{gardnerlaurentseries}).
We summarize the main steps of the Painlev\'{e} test below; refer to~\cite{baldwin-hereman-jnlmp-2006} \mbox{for details}.
\vskip 5pt
\noindent
{\bf Step 1}:
% MDPI: Please check if the bold and no indent format should be retained. 
% same as below.
% WH: Good. 
Compute the leading order term. 
To determine $\sigma$ and $u_0$, substitute $u_0 \, g^{\sigma}$
into (\ref{gardner}). 
Balance the minimal power terms in $g$, namely, $g^{3\sigma-1}$ and $g^{\sigma-3}$, 
to get $\sigma = -1$. 
Next, require that the leading terms (in $g^{-4})$ vanish, yielding
\begin{eqnarray}
\label{gardnerlaurentu0real}
u_0 &=& \pm \sqrt{-\tfrac{6}{\beta}} \, g_x \quad {\mathrm{if}} \; \beta < 0, 
\quad {\mathrm{or}} 
\\
\label{gardnerlaurentu0imaginary}
u_0 &=& \pm i \sqrt{\tfrac{6}{\beta}} \, g_x \;\;\quad {\mathrm{if}} \; \beta > 0. 
\end{eqnarray}
\vskip 2pt
\noindent
{\bf Step 2}: Compute the resonances.  
In this step, one determines which functions $u_k(x,t)$ in (\ref{gardnerlaurentseries}) 
will remain arbitrary. 
That happens at specific values of $k$, called {\em resonances}, 
denoted by $r$.
To find the resonances, substitute
$ u_0 \,  g^{\sigma} + u_r \, g^{\sigma + r}$, with~$\sigma = -1$ and $u_0$ 
given in (\ref{gardnerlaurentu0real})  or (\ref{gardnerlaurentu0imaginary}) 
into (\ref{gardner}), and~equate the coefficients of the dominant terms 
(in $g^{r-4}$) 
that are {\em linear} in $u_r$ to get the characteristic equation
\begin{equation}
\label{gardnerresonances}
(r-4)(r-3)(r+1) u_r g_x^3 = 0.
\end{equation}
Since
% MDPI: Please check if the no indent format should be retained. 
% same as below.
% WH: Good. 
$g_x \ne 0,$ the resonances are 
$r_1 = -1, r_2 = 3$, and~$r_3 = 4.$ 
Resonance $r_1 = -1$ corresponds to the arbitrariness of $g$.
\vskip 5pt
\noindent
{\bf Step 3}: Check the compatibility conditions.  
To do so, substitute
\begin{equation}
\label{gardnerlaurentseriestruncated}
u = \frac{1}{g} \sum_{k=0}^{4} u_k \, g^k, 
\end{equation}
into (\ref{gardner}) and verify that $u_1$ and $u_2$ can be {\em unambiguously} 
determined. 
Verify also that $u_3$ and $u_4$ are indeed arbitrary functions, 
meaning that no {\em compatibility conditions} arise. 
For (\ref{gardner}), one readily determines the real functions,
\begin{eqnarray}
\label{gardnerlaurentu1}
u_1 &=& -\frac{\alpha}{2\beta} 
             \pm \tfrac{1}{2} \sqrt{-\tfrac{6}{\beta}} 
             \left( \frac{g_{xx}}{g_x} \right), 
\\
\label{gardnerlaurentu2}
u_2 &=& \pm \left(
  \frac{4 \beta g_x g_t - \alpha^2 g_x^2 - 6 \beta g_{xx}^2 + 4 \beta g_x g_{3x}}
  {4 \beta \sqrt{-6 \beta}\, g_x^3}
  \right), 
\end{eqnarray}
when $\beta < 0$ (and complex expressions when $\beta > 0$).
At resonance $r = 3$, the~compatibility condition 
(arising as the coefficient of $g^{-3}$),
\begin{equation}
\label{gardnerresonance3}
 \Big( 
   u_0^2 \left( \alpha + 2 \beta u_1 \right) 
   - 6 \left( \, (u_0)_x g_x + u_0 g_{xx} \right) 
 \Big) g_x 
 - \beta u_0^2 (u_0)_x = 0,  
\end{equation}
is {\em identically} satisfied upon substitution of (\ref{gardnerlaurentu0real}) 
and (\ref{gardnerlaurentu1}). 
Likewise, a~long but straightforward computation shows that the compatibility 
condition at $r = 4$ (appearing at order $g^{-2}$), 
% MDPI: units should be normal format. please check and revise. 
% WH: What is normal format? 
\begin{eqnarray}
\label{gardnerresonance4}
 && 
 \Big( 
   u_0 u_1 \left( \alpha + \beta u_1 \right) 
   + \beta u_0^2 u_2 + 3 (u_0)_{xx}  
 \Big) g_x 
 + u_0 g_t - u_0 (u_0)_x \left(\alpha + 2 \beta u_1 \right)
 \nonumber \\
 && \; - \beta u_0^2 (u_1)_x + 3 (u_0)_x g_{xx} + u_0 g_{3x} = 0,  
\end{eqnarray}
is {\em identically} satisfied upon substitution of (\ref{gardnerlaurentu0real}), (\ref{gardnerlaurentu1}), and~(\ref{gardnerlaurentu2}).
  
Thus, at~least in the neighborhood of $g(x,t) = 0$, solution (\ref{gardnerlaurentseriestruncated}) is free of algebraic and logarithmic movable 
branch points. 
Apart from possible essential singularities (which the test is unable to detect), 
the movable singularities of its general solution are poles determined 
by $g(x,t) = 0$. 
Note that (\ref{gardnerlaurentseries}) serves as a {\em general} solution because, 
as required by the Cauchy--Kovalevskaya theorem, (\ref{gardner}) is of order $3$ 
in $x$ and that number agrees with $3$ arbitrary functions $g(x,t)$, 
$u_3(x,t),$ and $u_4(x,t)$ in (\ref{gardnerlaurentseries}).   

Moreover, notice that truncating the Laurent series at the constant level term yields 
an auto-B\"{a}cklund transformation,
\begin{equation}
\label{gardnertruncatedlaurentseries}
u(x,t) = \pm i \sqrt{\tfrac{6}{\beta}} \left(\tfrac{g_x}{g} \right) + u_1(x,t)
       = \pm i \sqrt{\tfrac{6}{\beta}} \left( \ln g \right)_x + u_1(x,t), 
\end{equation}
because for an arbitrary $g(x,t)$, both $u(x,t)$ and $u_1(x,t)$ must 
then be solutions of (\ref{gardner}) with $\beta > 0$.
Setting $u_1(x,t) = 0$ in (\ref{gardnertruncatedlaurentseries}) motivates the 
Hirota transformation discussed in Section~\ref{bilinear-form}, where it is shown
that only for $\beta > 0$ one can find soliton solutions of (\ref{gardner}).

In conclusion, the~Gardner equation passes the Painlev\'{e} test and, therefore, 
has the Painlev\'{e} property, which is a good predictor that the equation has  
conservation laws, generalized symmetries, and~so on. 
The investigation and actual computation of these quantities is based on 
a scaling symmetry of (\ref{gardner}), which we will discuss~next.

% SCALING SYMMETRY
\section{Scaling~Symmetry}
\label{scaling-symmetry}
As stated, when $\beta = 0$, the Gardner equation reduces to the KdV
Equation~(\ref{kdv}), which is scaling homogeneous under the scaling (dilation) symmetry
\begin{equation}
\label{kdvscale}
(x, t, u) \rightarrow 
(\kappa^{-1} x, \kappa^{-3} t, \kappa^2 u) 
= ( {\tilde{x}}, {\tilde{t}}, {\tilde{u}} ), 
\end{equation}
where $\kappa \ne 0$ is an arbitrary scaling parameter.
Indeed, replacing $(x, t, u)$ in the KdV equation in terms of 
$({\tilde{x}}, {\tilde{t}}, {\tilde{u}})$ 
yields
\begin{equation}
\label{kdvscalingtest}
\kappa^{-5} (
 {\tilde u}_{\tilde t} + \alpha {\tilde u} {\tilde u}_{\tilde x} 
 + {\tilde u}_{3{\tilde x}} ) = 0.
\end{equation}
A fast way to compute (\ref{kdvscale}) is to 
introduce the notions of weight, rank, and~uniformity of rank.
The {\em weight}, $W$, of~a variable is the exponent of $\kappa$ that 
multiplies that variable.
Thus, based on (\ref{kdvscale}), $W(x) = -1, W(t) = -3$, and~$W(u) = 2$.  
Equivalently, 
$W(\frac{\partial}{\partial x}) = 1$
and $W(\frac{\partial}{\partial t}) = 3.$
The {\em rank} of a monomial is defined as the total weight of the monomial.
For example, $\alpha u u_x$ has rank $5$ since the parameter $\alpha$ 
has no weight. 
An expression (or equation) is {\em uniform in rank} if all its monomial 
terms have equal~rank.

Now, if~(\ref{kdvscale}) were not known yet, it could quickly be found  
by requiring that the KdV equation is uniform in rank, yielding
\begin{equation}
\label{kdvweightequations} 
W(u) + W(\frac{\partial}{\partial t}) 
= 2 W(u) + W(\frac{\partial}{\partial x}) 
= W(u) + 3 W(\frac{\partial}{\partial x}).
\end{equation}
Solving (\ref{kdvweightequations}) gives 
$W(\frac{\partial}{\partial t}) = 3 W(\frac{\partial}{\partial x})$ and 
$ W(u) = 2 W(\frac{\partial}{\partial x}).$ 
Since $\kappa$ is arbitrary, without~loss of generality 
one can set $W(\frac{\partial}{\partial x}) = 1$, 
resulting in $W(\frac{\partial}{\partial t}) = 3$ and $W(u) = 2.$
Thus, requiring uniformity in rank of a PDE allows one to compute 
the weights of the variables (and, hence, the~scaling symmetry) 
with linear~algebra.

Setting $\alpha=0$, the~Gardner equation becomes the mKdV equation.
Requiring uniformity in rank readily yields 
$W(\frac{\partial}{\partial t}) = 3$ and $W(u) = 1.$
Hence, the~mKdV equation is scaling homogeneous under the transformation
\begin{equation}
\label{mkdvscale}
(x, t, u) \rightarrow 
(\kappa^{-1} x, \kappa^{-3} t, \kappa u). 
\end{equation}
Because $W(u)$ is different for the KdV and mKdV equations, 
the Gardner equation~(\ref{gardner}) will not be uniform in rank unless 
we use a trick. 
We will let the parameter $\alpha$ scale with some power of $\kappa$. 
Doing so, we must solve
\begin{equation}
\label{gardnerweightequations} 
W(u) + W(\frac{\partial}{\partial t}) 
= W(\alpha) + 2 W(u) + W(\frac{\partial}{\partial x}) 
= 3 W(u) + W(\frac{\partial}{\partial x}) 
= W(u) + 3 W(\frac{\partial}{\partial x}),
\end{equation}
yielding
\begin{equation}
\label{gardnerweights} 
W(\frac{\partial}{\partial x}) = 1, \;\; 
W(\frac{\partial}{\partial t}) = 3, \;\;
W(u) = 1, \;\; 
W(\alpha) = 1, 
\end{equation}
which expresses that (\ref{gardner}) has the scaling symmetry
\begin{equation}
\label{gardnerscale}
(x, t, u, \alpha) \rightarrow 
(\kappa^{-1} x, \kappa^{-3} t, \kappa u, \kappa \alpha).
\end{equation}
Starting with conservation laws, in~what follows, we will use
scaling symmetry to compute important quantities related to (\ref{gardner}), thereby establishing its complete~integrability. 

% CONSERVATION LAWS 
\section{Conservation~Laws}
\label{conservation-laws}
A {\em conservation law} of (\ref{scalarpde}) is an equation of the form
\begin{equation}
\label{conslawscalarpde} 
{\mathrm D}_{t} \, \rho + {\mathrm D}_{x} \, J \,\dot{=} \, 0,
\end{equation}
where the dot means that the equality should only hold on solutions 
$u(x,t)$ of (\ref{scalarpde}). 
$\rho$ is called a {\em conserved density} (or charge), and 
$J$ is the corresponding {\em flux} (or current). 

The scalar functions $\rho$ and $J$ are functions of $u$ and its partial derivatives
with respect to $x$.
All subsequent computations are performed on the jet space, which means that $u$ and 
its partial derivatives with respect to $x$ are treated as independent, 
alongside all monomials such as $u^3, u_x^2,$ etc. 
The density and flux could also explicitly depend on $x$ and $t$, 
but we will not cover such exceptional~cases.

In (\ref{conslawscalarpde}), 
${\mathrm D}_{t}$ is the total derivative with respect to $t,$ defined by
\begin{equation}
\label{gardnerDtrho} 
{\mathrm D}_t \, \rho 
= \frac{\partial \rho}{\partial t} + \rho^{\prime}[u_t] 
= \frac{\partial \rho}{\partial t} + 
\sum_{k=0}^{M} \frac{\partial \rho}{\partial u_{kx}} {\mathrm D}_x^k u_t,
\end{equation}
where $\rho^{\prime}[u_t]$ is the Fr\'echet derivative of $\rho$ in 
the direction of $u_t$, and $M$ is the highest order of $\rho$ in $x$.
In practice, one simply applies the chain rule for differentiation 
with respect to $t$, treating $u, u_x, u_{xx}$, etc., as~independent~functions.  

Likewise, ${\mathrm D}_{x}$ is the total derivative with respect to $x$,
\begin{equation}
\label{gardnerDxJ} 
{\mathrm D}_x J 
= \frac{\partial J}{\partial x} 
+ \sum_{k=0}^{N} \frac{\partial J}{\partial u_{kx}} {\mathrm D}_x (u_{kx})
= \frac{\partial J}{\partial x} 
+ \sum_{k=0}^{N} \frac{\partial J}{\partial u_{kx}} u_{(k+1)x},
\end{equation}
where $N$ is the order of $J$. 

Since (\ref{conslawscalarpde}) is {\em linear} in the densities (and fluxes),
a linear combination of densities with constant coefficients is still a density. 
The matching flux would, of~course, be a linear combination of the corresponding 
fluxes with the same constant~coefficients.

Returning to (\ref{gardner}), one can readily verify that
\begin{eqnarray}
\label{gardnerconslaw1}
&& 
{\mathrm D}_{t}(u) + {\mathrm D}_{x} 
(\tfrac{1}{2} \alpha u^2 + \tfrac{1}{3} \beta u^3 + u_{xx}) = 0, 
\\
\label{gardnerconslaw2} 
&& 
{\mathrm D}_{t}(u^2) + {\mathrm D}_{x} 
(\tfrac{2}{3} \alpha u^3 + \tfrac{1}{2} \beta u^4 - u_x^2 + 2 u u_{xx}) = 0, 
\\
\label{gardnerconslaw3} 
&& 
{\mathrm D}_{t}(\alpha u^3 + \tfrac{1}{2} \beta u^4 - 3 u_x^2) 
+ {\mathrm D}_{x} 
\left( \tfrac{3}{4} \alpha^2 u^4 + \alpha \beta u^5 + \tfrac{1}{3} \beta^2 u^6
- 6 \alpha u u_x^2 - 6 \beta u^2 u_x^2 \right.
\nonumber \\
&& 
\left. + 3 \alpha u^2 u_{xx} + 2 \beta u^3 u_{xx} + 3 u_{xx}^2 - 6 u_x u_{3x}
\right) = 0.
\end{eqnarray}
Indeed, (\ref{gardnerconslaw1}) is (\ref{gardner}) written as 
a conservation law. 
Next, (\ref{gardnerconslaw2}) straightforwardly follows after multiplication 
of (\ref{gardner}) by $2 u$ and a bit of integration by parts. 
% (to get the flux). 
Clearly, (\ref{gardnerconslaw3}) is far less obvious and will require
a computational strategy~\cite{goktas-hereman-jsc-1997,poole-hereman-jsc-2011}
and the use of codes like  
\verb|InvariantsSymmetries.m|
\cite{goktas-hereman-code-invarsym-1997} or
\verb|ConservationLawsMD.m| 
\cite{poole-hereman-code-conslawsmd-2009}.

Notice that the above densities are uniform in rank. 
Indeed, $\rho^{(1)} = u$ has rank $1$, $\rho^{(2)} = u^2$ has rank $2$,
and $\rho^{(3)} $ is of rank $4$. 
The corresponding fluxes are also uniform, with ranks $3, 4,$ and $6.$
As a matter of fact, the~entire conservation laws themselves are uniform in rank, 
with ranks $4, 5,$ and $7$, respectively.
This comes as no surprise because the defining Equation~(\ref{conslawscalarpde})
is only non-trivial if evaluated on solutions of the PDE, and therefore 
the densities, fluxes, and~conservation laws ``inherit'' (or adopt) 
the scaling homogeneity of the given PDE (and all its other continuous and 
discrete symmetries, for that matter). 

It turns out that the list of conservation laws of (\ref{gardner})
continues ad infinitum. 
The Gardner equation has infinitely many conservation laws, 
which is a clear indication that the PDE is completely~integrable.

Using the scaling symmetry (\ref{gardnerscale}), we will now show how 
to compute (\ref{gardnerconslaw3}), which is the shortest possible density 
of rank $4$, since it is free of any terms that could be moved into 
\mbox{the flux.}
\vskip 5pt
\noindent
{\bf Step 1}:
% MDPI: Please check if the bold and no indent format should be retained. 
% same as below.
% Yes, keep bold. Present format is good. 
% 
Construct a candidate density of rank $4$ as follows:
Make a list of all monomials in $u$ and $\alpha$ of rank $4$ or 
less,  
% (for the construction of densities that explicit depend 
% on $x$ and $t$ we refer to~\cite{poole-hereman-jsc-2011}),
% MDPI: footnote format is not allowed. we move it here. please confirm.
% WH: Better to put the content of the footnote in the text. 
% WH: We removed the text that was originally a footnote because it break the
% flow of the presentation.
% 
i.e., $\{\{ u^4, \alpha u^3, \alpha^2 u^2, \alpha^3 u, \alpha^4 \}$, 
$\{ u^3, \alpha u^2, \alpha^2 u, \alpha^3 \}$, 
$\{ u^2, \alpha u, \alpha^2 \}$, $\{ u, \alpha \}$, 
$ \{ 1 \} \}$. 
For the construction of candidate densities, the constant terms 
$\alpha^4, \alpha^3, \ldots, 1$ can be removed.  
Then, for~each monomial in that list, apply the correct number of $x$-derivatives 
so that the resulting term has exactly rank $4$.
The terms in the first sublist need no derivatives. 
Those in the second sublist need a single derivative.
The next set of terms need two derivatives, etc. 
For example, for~the first element in the third sublist,
${\mathrm D}_x^2 u^2 = 2 u_x^2 + 2 u u_{xx}.$ 
Obviously, if~we carry out partial integration, the~highest derivative term 
$u u_{xx}$ only differs from $u_x^2$ by the $x$-derivative of $\tfrac{1}{2} u^2$ 
and therefore can be ignored. 
Likewise, terms like $u^2 u_x$, $\alpha u u_x$, $\alpha^2 u_x$, 
$\alpha u_{xx},$ and $u_{3x}$ can be neglected because they are 
$x$-derivatives of single-term monomials ($\frac{1}{3} u^3$, 
$\tfrac{1}{2} \alpha u^2$, etc.).
There is no need to put terms like $u u_{xx}$, $u^2 u_x$ in the density because 
they can be moved to the~flux. 

Gather the resulting monomials after stripping off numerical factors and 
removing scalar multiples of single-term densities of lower rank 
(with regard to (\ref{gardnerconslaw1}) and (\ref{gardnerconslaw2}) these 
are $\alpha^3 u$ and $\alpha^2 u^2$). 
Finally, linearly combine the remaining terms with constant \mbox{coefficients, 
yielding}
\begin{equation}
\label{gardnercandidaterho3}
\rho = c_1 u^4 + c_2 \alpha u^3 + c_3 u_x^2, 
\end{equation}
which is of first order in $x$ ($M = 1$).
\vskip 5pt
\noindent
{\bf Step 2}: Compute the undetermined coefficients. 
Using (\ref{gardnerDtrho}), first compute
\begin{equation}
\label{gardnerDtrho3explicit} 
{\mathrm D}_t \, \rho 
= (4 c_1 u^3 {\mathrm I} + 3 \alpha c_2 u^2 {\mathrm I} 
  + 2 c_3 u_x {\mathrm D}_x) [u_t]
= (4 c_1 u^3 + 3 \alpha c_2 u^2) u_t + 2 c_3 u_x u_{xt}, 
\end{equation}
where ${\mathrm D}_x^0 = {\mathrm I}$ is the identity operator.  
Replace $u_t$ and $u_{xt} = u_{tx} $ from (\ref{gardner}) to get
\begin{eqnarray}  
\label{gardnerDtrho3evaluated} 
E &=& 
- \left( (4 c_1 u^3 + 3 \alpha c_2 u^2)
(\alpha u u_x + \beta u^2 u_x + u_{3x}) 
+ (2 c_3 u_x ) (\alpha u u_x + \beta u^2 u_x + u_{3x})_x \right)
\nonumber \\
 &=& - \left(
\alpha (4 c_1 + 3 \beta c_2) u^4 u_x + 4 \beta c_1 u^5 u_x + 4 c_1 u^3 u_{3x}
+ 3 \alpha^2 c_2 u^3 u_x  + 3 \alpha c_2 u^2 u_{3x}
\right. 
\nonumber \\
&& 
\left. 
+ 2 \alpha c_3 u_x^3 + 2 \alpha c_3 u u_x u_{xx} + 4 \beta c_3 u u_x^3 
+ 2 \beta c_3 u^2 u_x u_{xx} + 2 c_3 u_x u_{4x}
\right).
\end{eqnarray}
Next, find the constants $c_1, c_2,$ and $c_3$ so that $E = {\mathrm D}_t \rho$ 
matches $-{\mathrm D}_x J$ for some flux $J$ (to be computed in Step 3 below). 
Mathematically, this means that $E$ must be {\em exact}. 
The Euler operator (variational derivative),
\begin{eqnarray}
\label{eulerux} 
{\cal{L}}_{u(x)} 
= \sum_{k=0}^{K} (-{\mathrm D}_x)^k \frac{\partial }{\partial u_{kx} }
= \frac{\partial }{\partial u} 
  - {\mathrm D}_x \frac{\partial }{\partial u_x} 
  + {\mathrm D}_{x}^2 \frac{\partial }{\partial u_{xx}} 
  - {\mathrm D}_{x}^3 \frac{\partial }{\partial u_{3x}} 
  + {\mathrm D}_{x}^4 \frac{\partial }{\partial u_{4x}} + \ldots , 
\end{eqnarray}
allows one to test exactness
~\cite{hereman-etal-book-birkhauser-2005,hereman-etal-mcs-2007,hereman-etal-book-nova-2009,poole-hereman-jsc-2011}, 
where $K$ is the order of the expression the Euler operator is applied to. 
Therefore, $K = 4$ for $E$ in (\ref{gardnerDtrho3evaluated}).
Consequently, (\ref{eulerux}) will terminate after five terms.
$E$ will be exact if ${\cal{L}}_{u(x)} E \equiv 0$ on the jet space 
(treating $u, u_x, u_{xx}$, etc., and~also all monomials in such variables as independent). 
The computation of the terms in (\ref{eulerux}) involves nothing more than 
partial differentiations followed by (total) differentiations with 
respect to $x$.
Of a total of $30$ terms (not listed) 
% MDPI: "Data not shown” should be avoided. Please cite as a 
% reference/Supplementary Materials or please delete this phrase
% WH: This is not data. We are just no showing the 30 terms in the long 
% expression. Replaced not shown by not listed. If that is not acceptable 
% then remove "(not listed)".
% 
generated, many terms are canceled, and one is left with
\begin{equation}
\label{gardnereulerux} 
{\cal{L}}_{u(x)} E 
= - 4 (6 c_1 + \beta c_3) (u_x^3 + 3 u u_x u_{xx}) 
  - 6 \alpha (3 c_2 + c_3) u_x u_{xx} 
\end{equation},
which must vanish identically, 
yielding the linear system 
$6 c_1 + \beta c_3 = 0$ and $3 c_2 + c_3 = 0$ with solution
$c_1 = \tfrac{1}{2}\beta, c_2 = 1, c_3 = -3$.
Substitute these constants into (\ref{gardnercandidaterho3}) to obtain
\begin{equation}
\label{gardnerrho3}
\rho^{(3)} = \alpha u^3 + \tfrac{1}{2} \beta u^4 - 3 u_x^2, 
\end{equation}
the same expression as in (\ref{gardnerconslaw3}).
If one were only interested in the density, the~computation would finish here. 
To continue with the computation of the flux (in the next step), 
substitute the constants into (\ref{gardnerDtrho3evaluated}), yielding
\begin{eqnarray}
\label{gardnerDtrho3final} 
E &=& 
- \left(
 5 \alpha \beta u^4 u_x + 2 \beta^2 u^5 u_x + 2 \beta u^3 u_{3x} 
 + 3 \alpha^2 u^3 u_x + 3 \alpha u^2 u_{3x}
\right. 
\nonumber \\
&& 
\left. 
- 6 \alpha u_x^3 - 12 \beta u u_x^3 - 6 \alpha u u_x u_{xx} 
- 6 \beta u^2 u_x u_{xx} - 6 u_x u_{4x}
\right). 
\end{eqnarray}
 \vskip 2pt
\noindent
{\bf Step 3}: Compute the flux. 
Since ${\mathrm D}_x J^{(3)} = -{\mathrm D}_t \rho^{(3)} = -E$, to~get $J^{(3)}$ 
one must integrate $E$ with respect to $x$ and reverse the sign.  
There is a tool from differential geometry, called the {\em homotopy operator} 
\cite{olver-book-1993} p.\ 372, to~carry out integration by parts on the jet space. 
As will be shown below, application of the homotopy operator reduces the integration
on the jet space to a standard one-dimensional integration with respect to an 
auxiliary variable which will be denoted by $\lambda$. 

The homotopy operator for variable $u(x),$ acting on a exact expression $E$ of order $K$, is given~\cite{hereman-etal-book-birkhauser-2005,hereman-etal-mcs-2007,hereman-etal-book-nova-2009,poole-hereman-aa-2010,poole-hereman-jsc-2011}
by
\begin{equation}
\label{homotopyscalarux} 
{\cal H}_{u(x)} E = \int_{0}^{1} 
\left( I_u E \right) [\lambda u] \, \frac{d \lambda}{\lambda},
\end{equation}
with integrand
\begin{eqnarray}
\label{integrandhomotopyscalarux} 
I_u E 
& = & \sum_{k=1}^{K} 
\left( \sum_{i=0}^{k-1} u_{ix} (-{\mathrm D}_x)^{k-(i+1)} \right) 
\frac{\partial E}{\partial u_{kx}}
\nonumber \\
& = & ( u {\mathrm I})(\frac{\partial E}{\partial u_x}) 
+ (u_x {\mathrm I} - u {\mathrm D}_x) (\frac{\partial E}{\partial u_{xx}}) 
+ (u_{xx} {\mathrm I} - u_x {\mathrm D}_x + u {\mathrm D}_x^2) 
(\frac{\partial E}{\partial u_{3x}})
\nonumber \\
&& + ( u_{3x} {\mathrm I} - u_{xx} {\mathrm D}_x 
+ u_x {\mathrm D}_x^2 - u {\mathrm D}_x^3 )
(\frac{\partial E}{\partial u_{4x}}) + \ldots \, .
\end{eqnarray}
In (\ref{homotopyscalarux}), $(I_u E) [\lambda u]$ means that once 
$I_u E$ is computed one must replace $u$ by $\lambda u$, 
$u_x$ by $\lambda u_x $, $u_{xx}$ by $\lambda u_{xx}$, etc. 
Use (\ref{gardnerDtrho3final}), to~get
\begin{eqnarray}
\label{kdvIuforEpart2} 
{\mathrm I}_u E 
& = &
( u {\mathrm I} ) 
 (- 5 \alpha \beta u^4 - 2 \beta^2 u^5 - 3 \alpha^2 u^3
  + 18 \alpha u_x^2 + 36 \beta u u_x^2 + 6 \alpha u u_{xx}
  + 6 \beta u^2 u_{xx} + 6 u_{4x} ) 
\nonumber \\
&& + (u_x {\mathrm I} - u {\mathrm D}_x ) 
   (6 \alpha u u_x + 6 \beta  u^2 u_x)
  + ( u_{xx} {\mathrm I} - u_x {\mathrm D}_x + u {\mathrm D}_x^2 ) 
    (-3 \alpha u^2 -2 \beta u^3) 
\nonumber \\
&& + ( u_{3x} {\mathrm I} - u_{xx} {\mathrm D}_x + u_x {\mathrm D}_x^2 
   - u {\mathrm D}_x^3 ) (6 u_x)
\nonumber \\
& = & 
  - \left( 
  3 \alpha^2 u^4 + 5 \alpha \beta u^5 + 2 \beta^2 u^6 - 18 \alpha u u_x^2 
  - 24 \beta u^2 u_x^2 + 9 \alpha u^2 u_{xx} + 8 \beta u^3 u_{xx} 
\right.
\nonumber \\
&& \left. 
  + 6 u_{xx}^2 - 12 u_x u_{3x} 
  \right),
\end{eqnarray}
which already has the terms of $J^{(3)}$, but still with incorrect 
coefficients (and the opposite sign).
Finally, using (\ref{homotopyscalarux}),
\begin{eqnarray}
\label{gardnerJ3} 
J^{(3)} 
& = & - {\cal H}_{u(x)} (E) 
= - \int_0^1 (I_u (E))[\lambda u] \,\frac{d\lambda}{\lambda}
\nonumber \\
& = & 
\int_0^1 \left( 
3 \alpha^2 \lambda^3 u^4 + 5 \alpha \beta \lambda^4 u^5 + 2 \beta^2 \lambda^5 u^6
- 18 \alpha \lambda^2 u u_x^2 - 24 \beta \lambda^3 u^2 u_x^2 \right.
\nonumber \\
&& \left. 
+ 9 \alpha \lambda^2 u^2 u_{xx} + 8 \beta \lambda^3 u^3 u_{xx} 
+ 6 \lambda u_{xx}^2 - 12 \lambda u_x u_{3x}
\right) d\lambda
\nonumber \\
& = &
\tfrac{3}{4} \alpha^2 u^4 + \alpha \beta u^5 + \tfrac{1}{3} \beta^2 u^6
- 6 \alpha u u_x^2 - 6 \beta u^2 u_x^2 + 3 \alpha u^2 u_{xx} + 2 \beta u^3 u_{xx} 
\nonumber \\
&& 
+ 3 u_{xx}^2 - 6 u_x u_{3x},
\end{eqnarray}
which is exactly the flux in (\ref{gardnerconslaw3}).

We close this section with a remark about the use of conserved densities and constants of motion. 
If $J$ vanishes at infinity (because $u$ and its $x$-derivatives 
decay to zero), then 
integration of (\ref{conslawscalarpde}) with respect to $x$ yields
\begin{equation}
\label{conservedquantity} 
\frac{d}{dt} \int_{-\infty}^{\infty} \rho \, dx = 0.   
\end{equation}
Hence,
\begin{equation}
\label{constantmotion} 
P = \int_{-\infty}^{\infty} \rho \, dx
\end{equation}
is {\em constant} in time and often referred to as a conserved quantity or constant of motion. 
Depending on the physical setting, the~first few constants of motion 
express conservation of mass, momentum, and~energy.
Preserving these types of quantities plays an important role in testing the accuracy of numerical integrators. 
For example, some symmetric time-stepping methods~\cite{cohen-etal-nm-2008} 
and explicit finite-difference schemes~\cite{frasca-caccia-hydon-jcd-2019,sanz-serna-jcp-1982}
preserve two of these three conserved quantities of low degree,  
while the more general symplectic integrators preserve the Hamiltonian structure~\cite{ascher-mclachlan-jsc-2005,bridges-reich-jpa-2006}.
The reader is referred to, e.g.,~ \cite{hairer-etal-book-2006,leveque-book-1992,quispel-mclachlan-jpa-2006,sanz-serna-calvo-book-1994} 
for an in-depth discussion of this~subject.

% HIGHER-ORDER SYMMETRIES 
\section{Generalized~Symmetries}
\label{generalized-symmetries}
A second criterion to establish the complete integrability of (\ref{gardner})
is to show that the PDE has infinitely many generalized 
(higher-order) symmetries. 
As we will see, the~computation of a hierarchy of such symmetries~\cite{goktas-hereman-acm-1999} is algorithmic and easier than for 
conservation~laws. 

A scalar function $G(x, t, u, u_{x}, u_{xx}, \ldots, u_{mx})$ is called a 
{\em generalized symmetry} of (\ref{scalarpde}) 
if and only if it leaves (\ref{scalarpde}) invariant under the replacement 
$u \rightarrow u + \epsilon G$ within order $\epsilon$, that is,
\begin{equation}
\label{invariance}
{\mathrm D}_t (u + \epsilon G) \, \dot{=} \, F(u + \epsilon G)
\end{equation}
must hold up to order $\epsilon$ on the solutions of (\ref{scalarpde}). 
Consequently, $G$ must satisfy the {\em linearized} equation
\begin{equation}
\label{pdesymmetry}
{\mathrm D}_t G \, = \, \frac{\partial G}{\partial t} + F^{\prime}[G],
\end{equation}
where $F^{\prime}$ is the Fr\'{e}chet derivative of F in the direction 
of $G$:
\begin{eqnarray}
\label{pdefrechetG1}
F^{\prime}[G] 
&=& \frac{\partial}{\partial \epsilon} F(u + \epsilon G)_{|{\epsilon = 0}} 
\\
\label{pdefrechetG2}
&=& \sum_{k=0}^{n} \frac{\partial F}{\partial u_{kx}} 
{\mathrm D}_x^k G 
= \frac{\partial F}{\partial u} {\mathrm I}\, G 
+ \frac{\partial F}{\partial u_{x}} {\mathrm D}_x G
+ \frac{\partial F}{\partial u_{xx}} {\mathrm D}_x^2 G
+ \frac{\partial F}{\partial u_{3x}} {\mathrm D}_x^3 G
+ \ldots ,
\end{eqnarray}
where $n$ is the (highest) order of $F$.
In (\ref{invariance}) and (\ref{pdefrechetG1}), one must not only replace 
$u$ by $u + \epsilon G,$ but also $u_x$ by $u_x + \epsilon {\mathrm D}_x G$, \, 
$u_{xx}$ by $u_{xx} + \epsilon {\mathrm D}^2_x G$, etc. 
As before, ${\mathrm I}$ is the identity operator. 
The total derivative operators, ${\mathrm D}_t$ 
and ${\mathrm D}_x$, were defined in (\ref{gardnerDtrho}) 
and (\ref{gardnerDxJ}), respectively.
Since the defining Equation~(\ref{pdesymmetry}) is evaluated on solutions 
of the PDE, the~higher-order symmetries inherit the scaling homogeneity of (\ref{scalarpde}).

As we will show, (\ref{gardner}) has infinitely many generalized symmetries,
starting with
\begin{eqnarray}
\label{gardnersym1}
G^{(1)} &=& u_x, 
\\
\label{gardnersym2} 
G^{(2)} &=& \alpha u u_x + \beta u^2 u_x + u_{3x}, 
\\
\label{gardnersym3} 
G^{(3)} &=& 
\tfrac{5}{6} \alpha^2 u^2 u_x + \tfrac{5}{3} \alpha \beta u^3 u_x
+ \tfrac{5}{6} \beta^2 u^4 u_x + \tfrac{5}{3} \beta u_x^3 
+ \tfrac{10}{3} \alpha u_x u_{xx} + \tfrac{20}{3} \beta u u_x u_{xx}
\nonumber \\
&& + \tfrac{5}{3} \alpha u u_{3x} + \tfrac{5}{3} \beta u^2 u_{3x} + u_{5x}.
\end{eqnarray}
With regard to the weights in (\ref{gardnerweights}), the above symmetries 
are of ranks $2, 4$, and~$6$, respectively.  

Except for $G^{(1)}$, each of these symmetries leads to a completely 
integrable {\em nonlinear} PDE, 
$u_t + G^{(j)} = 0, \, j = 2, 3, \ldots $, 
where the weight of $t$ increases as $j$ increases. 
Notice that $u_t + G^{(2)} = 0$ corresponds to (\ref{gardner}) itself. 

Although the computation of symmetries is algorithmic~\cite{goktas-hereman-acm-1999}
and technically simpler than for densities, 
it still requires the computation of a plethora of terms, 
and is best 
handled with symbolic software such as 
\verb|InvariantsSymmetries.m|
\cite{goktas-hereman-acm-1999}.
As an example, we show how to compute $G^{(3)}$ of rank 6. 
\vskip 5pt
\noindent
{\bf Step 1}: Construct a candidate symmetry of rank $6$ as follows:
List all monomials in $u$ and $\alpha$ of rank $6$ or less, i.e.,~
$ \{ \{ u^6, \alpha u^5, \alpha^2 u^4, \alpha^3 u^3, \alpha^4 u^2, \alpha^5 u \}$, 
$ \{ u^5, \ldots, \alpha^4 u \}$, 
$ \{ u^4, \ldots, \alpha^3 u \}$, 
$ \{ u^3, \alpha u^2, \alpha^2 u \}$,
$ \{ u^2, \alpha u \} $, 
$ \{ u \} \} $, 
without the constant terms $\alpha^6, \alpha^5, \ldots, \alpha, 1$.  
\vskip 5pt
Then, apply the correct number of $x$-derivatives to the monomials in each of the 
six sublists so that the resulting terms have exactly rank $6$. 
The elements in the first sublist need no derivatives. 
Those in the second sublist need a single derivative, etc. 
For example, for~the first element in the fourth sublist, compute  
$\frac{\partial^3 u^3}{\partial{x^3}} 
  = 6 u_x^3 + 18 u u_x u_{xx} + 3 u^2 u_{3x}.$
Do this for each element in all sublists and gather the resulting monomials 
after stripping off numerical factors. 
To avoid lower-rank symmetries being recomputed, 
remove scalar multiples of single-term symmetries 
of lower rank, as well as scalar multiples of the highest-derivative term in
multiple-term symmetries of lower rank.
Thus, with~regard to (\ref{gardnersym1}) and (\ref{gardnersym2}), 
the monomials $\alpha^4 u_x$ and $\alpha^2 u_{3x}$ can be removed. 
Finally, linearly combine the remaining terms with constant coefficients, 
yielding
\begin{eqnarray}
\label{gardnercandidatersym3}
G & = & c_1 u^6 + \alpha c_2 u^5 + \alpha^2 c_3 u^4 
+ \alpha^3 c_4 u^3 + \alpha^4 c_5 u^2 + \alpha^5 c_6 u 
+ c_7 u^4 u_x + \alpha c_8 u^3 u_x 
\nonumber \\
&& + \alpha^2 c_9 u^2 u_x + \alpha^3 c_{10} u u_x
+ c_{11} u^2 u_x^2 + c_{12} u^3 u_{xx} + \alpha c_{13} u u_x^2 
+ \alpha c_{14} u^2 u_{xx} + \alpha^2 c_{15} u_x^2
\nonumber \\
&&  + \alpha^2 c_{16} u u_{xx} 
+ \alpha^3 c_{17} u_{xx} + c_{18} u_x^3 
+ c_{19} u u_x u_{xx} + c_{20} u^2 u_{3x} + \alpha c_{21} u_x u_{xx}
\nonumber \\
&&  + \alpha c_{22} u u_{3x} + c_{23} u_{xx}^2 + c_{24} u_x u_{3x} 
+ c_{25} u u_{4x} + \alpha c_{26} u_{4x} + c_{27} u_{5x} 
\end{eqnarray}
which is of fifth order ($m = 5$).
\vskip 5pt
\noindent
{\bf Step 2}: Find the undetermined coefficients. 
Compute ${\mathrm D}_t G$ and use (\ref{scalarpde}) to remove 
$u_t, u_{xt}$, etc. 
This produces an expression with 176 terms (not listed).
% MDPI: "Data not shown” should be avoided. Please cite as a 
% reference/Supplementary Materials or please delete this phrase
% WH: See my earlier remark. If "(not listed)" is not allowed then remove
% the "(not listed)".
Next, compute  (\ref{pdefrechetG2}) using
\begin{equation}
\label{gardnerF}
F = - ( \alpha u u_x + \beta u^2 u_x + u_{3x} )
\end{equation}
and $G$ from (\ref{gardnercandidatersym3}), yielding 193 terms 
(not listed). 
% MDPI: "Data not shown” should be avoided. 
% Please cite as a reference/Supplementary Materials 
% or please delete this phrase
% WH: See earlier remark. 
% 
Then, ${\mathrm D}_t G - F^{\prime}[G]$, which has 138 terms, must vanish identically on the jet space, yielding a linear system of
20 equations for the nine nonzero coefficients 
$(c_{7}, c_{8}, c_{9}$, $c_{18}$ through $c_{22}$, and~$c_{27})$:
\begin{eqnarray}
\label{gardnerGlinearsystem}
&& 2 (3 c_{20} - 5 \beta c_{27}) = 0, \,\, 
\alpha (3 c_{22} - 5 c_{27}) = 0, 
\nonumber \\
&& 
\phantom{,,\alpha (3 c_{22} - 5 c_{27}) = 0,} 
\vdots 
\phantom{\alpha (3 c_{22} - 5 c_{27}) = 0}
\nonumber \\
&& 3 \alpha (12 c_{8} - c_{19} - 2 \beta c_{21} - 4 \beta c_{22}) = 0, \,\, 
6 (3 c_{18} + 2 c_{19} + c_{20} - 20 \beta c_{27}) = 0.
\end{eqnarray}
For brevity, we have shown only a couple of the shortest equations 
(coming from $ u u_x u_{5x}$ and $u_x u_{5x}$, respectively) 
and two of the longest equations 
(coming from $ u u_x^2 u_{xx}$ and $u_x u_{xx} u_{3x}$, respectively).
The 18 coefficients $c_{1}$ through $c_{6}$, $c_{10}$ through $c_{17}$, 
and $c_{23}$ through $c_{26}$ are all zero.
Solving the system yields   
% the nonzero coefficients are
\begin{eqnarray} 
\label{gardnersolcGrank6}  
&& 
c_{7} = \tfrac{5}{6} \beta^2 c_{27}, \,\, 
c_{8} = \tfrac{5}{3} \beta c_{27}, \,\, 
c_{9} = \tfrac{5}{6} c_{27}, \,\, 
c_{18} = \tfrac{5}{3} \beta c_{27}, \,\,
c_{19} = \tfrac{20}{3} \beta c_{27}, \,\,
\nonumber \\
&& c_{20} = \tfrac{5}{3} \beta c_{27}, \,\, 
c_{21} = \tfrac{10}{3} c_{27}, \,\,
c_{22} = \tfrac{5}{3} c_{27}. 
\end{eqnarray}
Set $c_{27} = 1$ and substitute (\ref{gardnersolcGrank6}) into (\ref{gardnercandidatersym3}) to obtain
\begin{eqnarray}
\label{gardnersym3computed} 
G &=& 
\tfrac{5}{6} \alpha^2 u^2 u_x + \tfrac{5}{3} \alpha \beta u^3 u_x
+ \tfrac{5}{6} \beta^2 u^4 u_x + \tfrac{5}{3} \beta u_x^3 
+ \tfrac{10}{3} \alpha u_x u_{xx} + \tfrac{20}{3} \beta u u_x u_{xx}
\nonumber \\
&& + \tfrac{5}{3} \alpha u u_{3x} + \tfrac{5}{3} \beta u^2 u_{3x} + u_{5x}.
\end{eqnarray}
which matches $G^{(3)}$ in (\ref{gardnersym3}). 
 
The code \verb|InvariantsSymmetries.m| 
can also be used to verify if an evolution equation, 
e.g., $u_t + G^{(3)} = 0$, with~$G^{(3)}$ of rank 6,  
belongs to the hierarchy of completely integrable equations 
of a PDE of lower order. 
When asked to compute a symmetry of rank $4$ for 
$u_t + G^{(3)} = 0$, the~software returns $G^{(2)}$ in (\ref{gardnersym2}), 
confirming that $u_t + G^{(3)} = 0$ corresponds to a generalized symmetry of $u_t + G^{(2)} = 0$. 

% RECURSION OPERATOR 
\section{Recursion~Operator}
\label{recursion-operator}
To prove that there are infinitely many higher-order symmetries, 
we will compute the recursion operator, which generates those symmetries 
sequentially, starting from the lowest order symmetry $G^{(1)}$ in 
(\ref{gardnersym1}). 

As expected, the~recursion operator for the Gardner equation 
is a combination of the well-known recursion 
operators for the KdV and mKdV equations 
% WH: reformulated to get the correct numerical order of the references 
(see p.\ 312 of~\cite{olver-book-1993} and~\cite{olver-jmp-1977}):
% ~\cite{olver-jmp-1977} and~\cite{olver-book-1993} p. 312,
\begin{equation}
\label{gardnerR}
{\mathcal R} = 
{\mathrm D}_x^2 
+ \tfrac{2}{3} \left( \alpha u + \beta u^2 \right) {\mathrm I}
+ \tfrac{1}{3} \alpha u_x {\mathrm D}_x^{-1}
+ \tfrac{2}{3} \beta  u_x {\mathrm D}_x^{-1} (u {\mathrm I}), 
\end{equation}
where ${\mathrm D}_x^{-1}$ denotes the anti-derivative 
(or integral) operator.
${\mathcal R}$ connects the symmetries sequentially 
% MDPI:  Footnote format is not allowed. we move it here. please confirm.
% WH: FOOTNOTE issue resolved. Modified the text a little 
% to better incorporate the former footnote 
(without gaps in this case;
% the simplest cases; 
for examples with gaps, we refer to~\cite{baldwin-hereman-ijcm-2010}):
\begin{equation}
\label{gardnerlinkGjGj1}
G^{(j+1)} = {\mathcal R} G^{(j)}, \quad j = 1, 2, \ldots  . 
\end{equation}
For example,
\begin{eqnarray}
\label{gardnerlinkG1G2}
{\mathcal R} u_x 
&=& 
\left(
{\mathrm D}_x^2 
+ \tfrac{2}{3} \left( \alpha u + \beta u^2 \right) {\mathrm I}
+ \tfrac{1}{3} \alpha u_x {\mathrm D}_x^{-1}
+ \tfrac{2}{3} \beta u_x {\mathrm D}_x^{-1} (u {\mathrm I}) 
\right) 
u_x
\nonumber \\
&=& 
u_{3x} 
+ \tfrac{2}{3} \left( \alpha u + \beta u^2 \right) u_x
+ \tfrac{1}{3} \alpha u_x {\mathrm D}_x^{-1} (u_x)
+ \tfrac{2}{3} \beta u_x {\mathrm D}_x^{-1} (\tfrac{1}{2} u^2)_x
\nonumber \\
&=& \alpha u u_x + \beta u^2 u_x + u_{3x}, 
\end{eqnarray}
which is $G^{(2)}$, and
% WH: EQUATION POSITION. "Substraction of space" has been removed. 
% ~\vspace{-12pt}
% \begin{adjustwidth}{-\extralength}{0cm}
% WH: EQUATION POSITION. "centering and adjustwidth" removed 
% to avoid that equation runs over the left margin. 
% \centering 
%% If there is a figure in wide page, please release command \centering
% WH: EQUATION POSITION. Equation ran over left margin. 
% Format commands have been removed.  
% WH: EQUATION POSITION. Retracted small spaces below 
\begin{eqnarray}
\label{gardnerlinkG2toG3}
{\mathcal R} G^{(2)} 
\!\!&=&\!\! \left(
{\mathrm D}_x^2 \!+\! 
\tfrac{2}{3} \left( \alpha u \!+\! \beta u^2 \right) {\mathrm I} 
\!+\! \tfrac{1}{3} \alpha u_x {\mathrm D}_x^{-1}
\!+\! \tfrac{2}{3} \beta u_x {\mathrm D}_x^{-1} (u {\mathrm I}) 
\right) 
\left(\alpha u u_x \!+\! \beta u^2 u_x \!+\! u_{3x} \right)
\nonumber \\
\!\!&=&\!\! u_{5x} + \tfrac{5}{3} \alpha u u_{3x} 
+ \tfrac{5}{3} \beta u^2 u_{3x}
+ \ldots  
+ \tfrac{\alpha}{3} u_x {\mathrm D}_x^{-1} 
\left( \tfrac{1}{2} \alpha u^2 + \tfrac{1}{3} \beta u^3 
       + u_{xx} \right)_x
\nonumber \\
\!\!&&\!\! + \tfrac{2}{3} \beta u_x {\mathrm D}_x^{-1}
\left(
\tfrac{1}{3} \alpha u^3 + \tfrac{1}{4} \beta u^4 + u u_{xx} 
-\tfrac{1}{2} u_x^2
\right)_x 
\nonumber \\
\!\!&=&\!\! \tfrac{5}{6} \alpha^2 u^2 u_x + \tfrac{5}{3} \alpha \beta u^3 u_x
+ \tfrac{5}{6} \beta^2 u^4 u_x + \tfrac{5}{3} \beta u_x^3 
+ \tfrac{10}{3} \alpha u_x u_{xx} + \tfrac{20}{3} \beta u u_x u_{xx}
\nonumber \\
\!\!&&\!\! + \tfrac{5}{3} \alpha u u_{3x} 
+ \tfrac{5}{3} \beta u^2 u_{3x} + u_{5x}, 
\end{eqnarray}
% WH: EQUATION POSITION. "adjustwidth" removed to avoid that equation 
% runs over left margin
% \end{adjustwidth}
which is $G^{(3)}$.

Even for scalar equations like (\ref{gardner}), the~computation 
of recursion operators~\cite{baldwin-hereman-ijcm-2010} is lengthy, 
but can be performed with specialized software such as 
\verb|PDERecursionOperator.m|
\cite{baldwin-hereman-code-recoper-2003}.
One can also use computer algebra to verify that ${\mathcal R}$
 is a hereditary operator~\cite{fuchssteiner-etal-cpc-1987}. 

If ${\mathcal R}$ is a recursion operator for (\ref{scalarpde}), 
then the Lie derivative~\cite{hereman-goktas-wester-1999,olver-book-1993,wang-phdthesis-1998,wang-jnlmathphys-2002} 
of ${\mathcal R}$ is zero, yielding
\begin{equation}
\label{definingeqR}
{\mathrm D}_t {\mathcal R} + [ {\mathcal R}, {F}^{\prime}] 
= \frac{\partial {\mathcal R}}{\partial t} 
  + {\mathcal R}^{\prime}[F]
  + {\mathcal R} \circ F^{\prime} 
  - F^{\prime} \circ {\mathcal R} \, \dot{=} \, 0, 
\end{equation}
where $[ \, , \, ]$ and $\circ$ denote the commutator and composition 
of operators, respectively.
$F^{\prime}$ is the Fr\'echet derivative operator, defined as
\begin{equation}
\label{frechetF}
F^{\prime} 
= \sum_{k=0}^n \frac{\partial F}{\partial u_{kx}} {\mathrm D}_x^k,
\end{equation}
where $n$ is the order of $F$ in (\ref{scalarpde}). 
${\mathcal R}^{\prime}[F]$ is the Fr\'echet derivative of ${\mathcal R}$ 
in the direction of $F$. 
For recursion operators of the form
\begin{equation}
\label{formR}
{\mathcal R} 
 = \sum_{j=1}^T \, U_j(u, u_x, u_{xx}, \ldots, u_{m_1\, x}) \,
 {\mathcal S}_j 
 ( {\mathrm D}_x, {\mathrm D}_x^2, \ldots, {\mathrm D}_x^{-1} ) 
 \, ( V_j(u, u_x, u_{xx}, \ldots, u_{m_2 \,x})\, {\mathrm I} ), 
\end{equation}
where $T$ is the total number of terms in ${\mathcal R}$, and~
$m_1$ and $m_2$ are the orders of $U$ and $V$, respectively, 
one has
\begin{equation}
\label{frechetR}
{\mathcal R}^{\prime}[F] 
= \sum_{j=0}^{m_1} 
({\mathrm D}_x^{j} F) \, \frac{\partial U_j}{\partial u_{jx}} \,
{\mathcal S}_j \,(V_j \, {\mathrm I})
 + \sum_{j=0}^{m_2} 
 U_j \, {\mathcal S}_j \,
 \left( 
 ({\mathrm D}_x^{j} F) \,\frac{\partial V_j}{\partial u_{jx}}\, {\mathrm I} 
 \right).
\end{equation}
With regard to (\ref{gardnerR}), 
${\mathcal R}$ is a linear integro-differential operator~\cite{sanders-wang-nlanal-2001,sanders-wang-physicad-2001}
which naturally splits into two pieces~\cite{baldwin-etal-crm-2005,baldwin-hereman-ijcm-2010,wang-jnlmathphys-2002},
\begin{equation}
\label{recoperR0R1}
{\mathcal R} = {\mathcal R}_0 + {\mathcal R}_1,
\end{equation}
where ${\mathcal R}_0$ is a local differential operator 
and ${\mathcal R}_1$ is a non-local integral operator.  
${\mathcal R}_0$ is a linear combination of 
monomials involving ${\mathrm D}_x$, $u$, and~parameters with weight 
(if applicable) so that each monomial has the correct rank.
Note also that ${\mathrm D}_x$ will always be ``propagated'' to the right
in ${\mathcal R}_0$.
For example,
\begin{equation}
\label{Dx2onuI}
{\mathrm D}_x^2 (u {\mathrm I})  
= {\mathrm D}_x \left( u_x {\mathrm I} + u {\mathrm D}_x \right)
= u_{xx} {\mathrm I} + 2 u_x {\mathrm D}_x + u {\mathrm D}_x^2.
\end{equation}
Based on the theory of recursion operators, 
${\mathcal R}_1$ is a linear combination with constant coefficients 
of terms of the form
\begin{equation}
\label{structureR1}
  G^{(i)} {\mathrm D}_x^{-1} {\mathcal L}_u (\rho^{(j)}), 
\end{equation}
where $G^{(i)}$ is a symmetry and ${\mathcal L}_u(\rho^{(j)})$ 
is a cosymmetry (Euler operator applied to a density $\rho^{(j)})$, 
selected such that ${\mathcal R}_1$ has the correct rank~\cite{bilge-pha-1993,wang-jnlmathphys-2002}. 
To standardize the form of ${\mathcal R}_1$, one propagates 
${\mathrm D}_x$ to the left. 
For example, 
$ {\mathrm D}_x^{-1} u_x {\mathrm D}_x 
= u_x {\mathrm I} - {\mathrm D}_x^{-1} u_{xx} {\mathrm I}$.
The local and non-local operators are then added to obtain 
a candidate recursion~operator.  

We will now show how (\ref{gardnerR}) is computed. 
Since the ranks $G^{(1)}$ and $G^{(2)}$ (as well as $G^{(2)}$ and $G^{(3)}$) 
differ by $2$, ${\cal R}$ must have rank $2$. 
Based on (\ref{gardnerweights}), $W({\mathrm D}_x^{-1}) = -1$,
and one can readily verify that all the terms in (\ref{gardnerR}) have rank $2$.
\vskip 5pt
\noindent
{\bf Step 1}: Compute the candidate recursion operator. 
Using (\ref{gardnerweights}), list all monomials in 
${\mathrm D}_x, u$, and~$\alpha$ of rank $2$ or less, i.e.,~
$ \{ 
  \{ {\mathrm D}_x^2, u {\mathrm D}_x, \alpha {\mathrm D}_x$, 
  $u^2 {\mathrm I}, \alpha u {\mathrm I} \}$,
  $ \{ {\mathrm D}_x, u {\mathrm I} \}  
  \} $, 
where the trivial terms $\alpha {\mathrm I}$ and $\alpha^2 {\mathrm I}$
have been removed.  
Apply the correct number of $x$-derivatives to the monomials 
in each of the two sublists to assure that each term has rank $2$. 
No action is needed on the first sublist. 
The elements in the second sublist need a single derivative, yielding
$\{ {\mathrm D}_x^2, u_x {\mathrm I}, u {\mathrm D}_x \}$.
After stripping off numerical factors and removing duplicates,
linearly combine the resulting monomials with constant coefficients
to obtain a candidate local operator
\begin{equation}
\label{gardnerR0}
{\mathcal R}_0 
= c_1 {\mathrm D}_x^2 + c_2 u {\mathrm D}_x + \alpha c_3 {\mathrm D}_x
  + c_4 u^2 {\mathrm I} + \alpha c_5 u {\mathrm I}
  + c_6 u_x {\mathrm I}. 
\end{equation}
Using symmetry $G^{(1)} = u_x$ and densities $\rho^{(1)} = u$
and $\rho^{(2)} = u^2$ (all of low rank),   
compute ${\mathcal L}_u \rho^{(1)} = 1$ and ${\mathcal L}_u \rho^{(2)} = 2 u$.
Then, with~terms of type (\ref{structureR1}), make the candidate 
non-local operator
\begin{equation}
\label{gardnerR1}
{\mathcal R}_1 
= \alpha c_7 u_x {\mathrm D}_x^{-1} 
+ c_8 u_x {\mathrm D}_x^{-1} (u {\mathrm I}),
\end{equation}
so that each term has rank $2$.
Add both pieces to get the candidate recursion operator
\begin{equation}
\label{gardnercandidateR}
{\mathcal R} 
= c_1 {\mathrm D}_x^2 + ( c_2 u + \alpha c_3 ) {\mathrm D}_x
 + (c_4 u^2 + \alpha c_5 u + c_6 u_x) {\mathrm I} 
 + \alpha c_7 u_x {\mathrm D}_x^{-1} 
 + c_8 u_x {\mathrm D}_x^{-1} (u {\mathrm I}).
\end{equation}
\vskip 5pt
\noindent
{\bf Step 2}: Compute the undetermined coefficients. 
Separately compute all the pieces in (\ref{definingeqR}), 
beginning with
\begin{eqnarray}
\label{gardnerDtR}
{\mathrm D}_t {\mathcal R} 
&=& 
  c_2 u_t {\mathrm D}_x 
  + ( 2 c_4 u u_t + \alpha c_5 u_t + c_6 u_{xt} ) {\mathrm I}
  + \alpha c_7 u_{xt} {\mathrm D}_x^{-1} 
\nonumber \\
&&   
  + c_8 u_{xt} {\mathrm D}_x^{-1} (u {\mathrm I}) 
  + c_8 u_x {\mathrm D}_x^{-1} (u_t \, {\mathrm I}) 
\nonumber \\
\label{gardnerDtRonF}
&=& c_2 F {\mathrm D}_x 
   + ( 2 c_4 u F + \alpha c_5 F + c_6 {\mathrm D}_x F ) {\mathrm I}
   + \alpha c_7 ({\mathrm D}_x F) {\mathrm D}_x^{-1} 
\nonumber \\
&& 
  + c_8 ({\mathrm D}_x F) {\mathrm D}_x^{-1} (u {\mathrm I}) 
  + c_8 u_x {\mathrm D}_x^{-1} (F \,{\mathrm I}), 
\end{eqnarray}
which can also be computed using (\ref{frechetR}).
With $F$ given in (\ref{gardnerF}), insert
\begin{equation}
\label{gardnerFx}
{\mathrm D}_x F = F_x 
= -( \alpha u_x^2 + \alpha u u_{xx} + 2 \beta u u_x^2 + \beta u^2 u_{xx} + u_{4x} )  
\end{equation}
into (\ref{gardnerDtR}), expand, and simplify. 
This yields an expression of 27 terms 
(not listed).
% MDPI: "Data not shown” should be avoided. Please cite as a 
% reference/Supplementary Materials or please delete this phrase
% WH: See earlier remark.
Some of the terms have ${\mathrm D}_x$ and ${\mathrm I}$; 
others involve 
${\mathrm D}_x^{-1}$, 
${\mathrm D}_x^{-1} (u {\mathrm I})$, 
${\mathrm D}_x^{-1} (u u_x {\mathrm I})$,
${\mathrm D}_x^{-1} (u^2 u_x {\mathrm I})$, 
and 
${\mathrm D}_x^{-1} (u_{3x} {\mathrm I})$.  
Next, using (\ref{frechetF}), compute
\begin{equation}  
\label{gardnerfrechetF}
F^{\prime} = 
 -\left( 
 {\mathrm D}_x^{3} + (\alpha + \beta u) u {\mathrm D}_x 
 + (\alpha + 2 \beta u) u_x {\mathrm I}
\right).
\end{equation}
With (\ref{gardnercandidateR}) and (\ref{gardnerfrechetF}), 
then compute
\begin{eqnarray}
\label{gardnerRcircFprime}
{\mathcal R} \circ F^{\prime}
&=& 
- \left( 
    \left( c_1 {\mathrm D}_x^2 + (c_2 u + \alpha c_3) {\mathrm D}_x
       + (c_4 u^2 + \alpha c_5 u + c_6 u_x) {\mathrm I} 
       + \alpha c_7 u_x {\mathrm D}_x^{-1} 
    \right. 
  \right.
\nonumber \\
&& 
\left. 
  \left. 
    \quad\;\; 
    + c_8 u_x {\mathrm D}_x^{-1} (u {\mathrm I}) 
  \right) \circ 
% circ
  \left( 
  {\mathrm D}_x^{3} + (\alpha + \beta u) u {\mathrm D}_x 
  + (\alpha + 2 \beta u) u_x {\mathrm I} 
  \right)
\right), 
\end{eqnarray}
using, for~example,
\begin{eqnarray}
\label{gardnerDxTwousqDx}
{\mathrm D}_x^2 ( u^2 {\mathrm D}_x ) 
&=& {\mathrm D}_x ( 2 u u_x {\mathrm D}_x + u^2 {\mathrm D}_x^2 )
\nonumber \\
&=& 2 u_x^2 {\mathrm D}_x + 2 u u_{xx} {\mathrm D}_x 
    + 4 u u_x {\mathrm D}_x^2 + u^2 {\mathrm D}_x^3,
\end{eqnarray}
to consistently move the operator ${\mathrm D}_x$ 
from the left to the right.  
To be able to match the integral terms in (\ref{gardnerDtR}) 
which are mentioned below (\ref{gardnerFx}),
repeated integration by parts is needed~\cite{baldwin-hereman-ijcm-2010}. 
For example,
\begin{eqnarray}
\label{intbyparts}
{\mathrm D}_x^{-1} (u {\mathrm D}_x^3)
&=& u {\mathrm D}_x^2 
    - {\mathrm D}_x^{-1} (u_x {\mathrm D}_x^2)
\nonumber \\
&=& u {\mathrm D}_x^2 
    - u_x {\mathrm D}_x 
    + {\mathrm D}_x^{-1} (u_{xx} {\mathrm D}_x)
\nonumber \\    
&=& u {\mathrm D}_x^2 
    - u_x {\mathrm D}_x 
    + u_{xx} {\mathrm I} 
    - {\mathrm D}_x^{-1} (u_{3x} {\mathrm I}),
\end{eqnarray}
converts the integral ${\mathrm D}_x^{-1} (u {\mathrm D}_x^3)$
into ${\mathrm D}_x^{-1} (u_{3x} {\mathrm I})$ by moving
the ${\mathrm D}_x$ operator under the ${\mathrm D}_x^{-1}$ 
operator from the right to the left. 
After integration by parts, ${\mathcal R} \circ F^{\prime}$ has
58 terms (not listed). 
% MDPI: "Data not shown” should be avoided. Please cite as a 
% reference/Supplementary Materials or please delete this phrase
% WH: See earlier remark.
% 
Next, compute
\begin{eqnarray}
\label{gardnerFprimecircR}
F^{\prime} \circ {\mathcal R}
&\!=\!& 
- \left( 
    \left(
    {\mathrm D}_x^{3} + (\alpha + \beta u) u {\mathrm D}_x 
    + (\alpha + 2 \beta u) u_x {\mathrm I} 
    \right) \circ 
   \left(
  c_1 {\mathrm D}_x^2 + (c_2 u + \alpha c_3) {\mathrm D}_x
  \right.
  \right.
\nonumber \\
&& 
\left. 
  \left. 
  \quad \;\;
  + (c_4 u^2 + \alpha c_5 u + c_6 u_x) {\mathrm I} 
  + \alpha c_7 u_x {\mathrm D}_x^{-1}  
  + c_8 u_x {\mathrm D}_x^{-1} (u {\mathrm I}) 
  \right)
\right). 
\end{eqnarray}
To move the operator ${\mathrm D}_x$ to the right, use, for~example,
\begin{eqnarray}
\label{gardnerDxcubeduxIntDxuI}
\!\!\!{\mathrm D}_x^3 \left( u_x {\mathrm D}_x^{-1} (u {\mathrm I}) \right)
\!&\!=\!&\! 
{\mathrm D}_x^2 
  \left( 
  u_{xx} {\mathrm D}_x^{-1}(u {\mathrm I}) + u u_x {\mathrm I} 
  \right)
\nonumber \\
\!&\!=\!&\!
{\mathrm D}_x 
  \left( 
  u_{3x} {\mathrm D}_x^{-1}(u {\mathrm I}) + 2 u u_{xx} {\mathrm I} 
  + u_x^2 {\mathrm I} + u u_x {\mathrm D}_x
  \right)
\nonumber \\
\!&\!=\!&\! u_{4x} {\mathrm D}_x^{-1} (u {\mathrm I}) 
    +\! 4 u_x u_{xx} {\mathrm I} 
    +\! 3 u u_{3x} {\mathrm I} 
    +\! 3 u u_{xx} {\mathrm D}_x 
    +\! 2 u_x^2 {\mathrm D}_x 
    +\! u u_x {\mathrm D}_x^2, 
\end{eqnarray}
and similar formulas 
(expressed in a more general form in~\cite{baldwin-hereman-ijcm-2010}). 
The expanded expression for $F^{\prime} \circ {\mathcal R}$ 
has 73 terms. 
Substitute the simplified expressions for (\ref{gardnerDtR}), 
(\ref{gardnerRcircFprime}), and~(\ref{gardnerFprimecircR}) into 
(\ref{definingeqR}) and require that the resulting expression
(which has 36 terms) vanishes identically, i.e.,~all monomials in 
$u, u_x, \ldots, {\mathrm I}$, ${\mathrm D}_x, 
{\mathrm D}_x^2$, $\dots$,
should be treated as independent.  
This yields 
$c_{2} = c_{3} = c_{6} = 0$ and a linear system,  
% GOOD VERIFIED
\begin{eqnarray}
\label{gardnerRlinearsystem}
&& 2 (3 c_{4} - 2 \beta c_{1}) = 0, \,\, 
\alpha (3 c_{5} - 2 c_{1}) = 0, \,\, 
\alpha (3 c_{7} - c_{1}) = 0, \,\, 
3 c_{8} - 2 \beta c_1 = 0, \,\, 
\nonumber \\
&& 
3 \alpha (c_{7} + c_{5} - c_{1}) = 0, \,\,
3 (c_{8} + 2 c_{4} - 2 \beta c_1 ) = 0,  
\end{eqnarray}
involving the remaining nonzero constants. 
Solve the system to get 
% GOOD VERIFIED
\begin{equation} 
\label{gardnersolcR} 
c_{4} = \tfrac{2}{3} \beta c_{1}, \,\, 
c_{5} = \tfrac{2}{3} c_{1}, \,\, 
c_{7} = \tfrac{1}{3} c_{1}, \,\, 
c_{8} = \tfrac{2}{3} \beta c_{1}. 
\end{equation}
Finally, set $c_{1} = 1$ and substitute (\ref{gardnersolcR}) into 
(\ref{gardnercandidateR}), yielding
\begin{equation}
\label{gardnerRcomputed}
{\mathcal R} = 
{\mathrm D}_x^2 
+ \tfrac{2}{3} \left( \alpha u + \beta u^2 \right) {\mathrm I}
+ \tfrac{1}{3} \alpha u_x {\mathrm D}_x^{-1}
+ \tfrac{2}{3} \beta u_x {\mathrm D}_x^{-1} (u {\mathrm I}), 
\end{equation}
which matches ${\mathcal R}$ in (\ref{gardnerR}). 

% LAX PAIR
\section{Lax~Pair}
\label{lax-pair}
Another way to prove the complete integrability of a nonlinear PDE 
of type (\ref{scalarpde}) is to construct a Lax pair consisting 
of two {\em linear} PDEs in an auxiliary function whose compatibility 
requires that the original PDE is~satisfied.

There are two flavors of Lax pairs.
One is an operator formulation where the Lax pair consists of a pair of differential 
operators which leads to a higher order linear equation involving an auxiliary function.
Alternatively, in~the matrix formulation, the Lax pair is a set of two matrices 
satisfying a system of equations of first order in $x$ and $t$, respectively. 
Only Lax pairs in operator form will be covered in this section. 
The reader is referred to the literature~\cite{ablowitz-clarkson-book-1991,ablowitz-segur-book-1981,hickman-etal-aa-2012}
for the matrix formalism (also called the zero curvature representation). 

Finding a Lax pair in operator form is a nontrivial task and requires an 
educated guess about the order of the differential operators.  
But once the order is selected, one can take advantage of the 
scaling symmetry of the given PDE to construct a candidate for each operator because they inherit that scaling symmetry.
Here again, the~defining equation for a Lax pair is evaluated on solutions 
of the given PDE which supports that~claim.

Using the KdV equation as prime example, Lax~\cite{lax-cpam-1968} 
showed that a completely integrable {\em nonlinear} PDE has an associated 
system of {\em linear} PDEs involving a pair of linear differential operators 
$({\mathcal L},\, {\mathcal M})$ and an auxiliary function $\psi(x,t)$,
\begin{equation}
\label{LaxLandM}
{\mathcal L} \psi = \lambda \psi, \;\; \psi_t = {\mathcal M} \psi, 
\end{equation}
where ${\mathcal L}$ and ${\mathcal M}$ are expressed in powers of 
${\mathrm D}_x$ with coefficients depending on $u, u_x$, etc., and~
$\psi$ is an eigenfunction of ${\mathcal L}$ corresponding 
to eigenvalue $\lambda$.
To guarantee complete integrability of (\ref{scalarpde}), 
at least one non-trivial Lax pair 
$({\mathcal L},\, {\mathcal M})$ should exist, 
and the eigenvalue should not change in time which makes the problem 
{\em isospectral}.

We will show below that a one-parameter family of Lax pairs of (\ref{gardner}) 
is given by
\begin{eqnarray}
\label{gardnerLresult}
{\mathcal L} &=&  
 {\mathrm D}_x^2 
 + 2 \epsilon u \, {\mathrm D}_x 
 + \tfrac{1}{6} \left( 
    (6 \epsilon^2 + \beta) u^2 
    + \alpha u  
    + (6 \epsilon \pm \sqrt{-6\beta}) u_x 
 \right) \, {\mathrm I}, 
\\
\label{gardnerMresult}
{\mathcal M} &=& 
  - 4 {\mathrm D}_x^3 
  - 12 \epsilon u \, {\mathrm D}_x^2  
  - \left(
    (12 \epsilon^2 + \beta ) u^2 
    + \alpha u  
    + (12 \epsilon \pm \sqrt{-6\beta}) u_x 
  \right) \, {\mathrm D}_x
\nonumber \\ 
&& 
-\left( 
 \tfrac{1}{3} \epsilon (12 \epsilon^2 + 2 \beta) u^3 
 + \tfrac{1}{2} \alpha \epsilon u^2  
 + (12 \epsilon^2 + \beta \pm \epsilon \sqrt{-6\beta}) u u_x 
 \right. 
\nonumber \\
&&
  \left. + \tfrac{1}{2} \alpha u_x  
  + \tfrac{1}{2} (6 \epsilon \pm \sqrt{-6\beta}) u_{xx}
 \right) \, {\mathrm I}, 
 \end{eqnarray}  
where $\epsilon$ is any real or complex number (not necessarily small).
Substituting ${\mathcal L}$ and ${\mathcal M}$ into (\ref{LaxLandM})
yields
\begin{eqnarray}
\label{gardnerLpsi}
{\mathrm D}_x^2 \psi 
&=& 
 - 2 \epsilon u \, {\mathrm D}_x \psi
 + \left( \lambda 
    - (\epsilon^2 + \tfrac{1}{6} \beta) u^2 
    - \tfrac{1}{6} \alpha u  
    - (\epsilon \pm \tfrac{1}{6} \sqrt{-6 \beta}) u_x  
 \right)\, \psi,
\\
\label{gardnerMpsi}
{\mathrm D}_t \psi 
&=&
 - 4 {\mathrm D}_x^3 \psi 
 - 12 \epsilon u \, {\mathrm D}_x^2 \psi 
 - \left(
    (12 \epsilon^2 + \beta ) u^2 
    + \alpha u  
    + (12 \epsilon \pm \sqrt{-6\beta}) u_x 
  \right) \, {\mathrm D}_x \psi
\nonumber \\ 
&& 
- \left( 
  \tfrac{1}{3} \epsilon (12 \epsilon^2 + 2 \beta) u^3 
   + \tfrac{1}{2} \alpha \epsilon u^2  
   + (12 \epsilon^2 + \beta \pm \epsilon \sqrt{-6\beta}) u u_x 
 \right. 
\nonumber \\
&&
  \left. 
   + \, \tfrac{1}{2} \alpha u_x  
   + \tfrac{1}{2} (6 \epsilon \pm \sqrt{-6\beta}) u_{xx}
  \right) \, \psi.
\end{eqnarray}
The first equation is a Schr\"{o}dinger-type equation for the arbitrary 
eigenfunction $\psi$ with eigenvalue $\lambda$ and potential 
$ (\epsilon^2 + \tfrac{1}{6} \beta) u^2 $
$ + \tfrac{1}{6} \alpha u $ 
$ + (\epsilon \pm \tfrac{1}{6} \sqrt{-6 \beta}) u_x$.
The second equation describes the time evolution of the~eigenfunction.

A lengthy computation shows that the compatibility condition for (\ref{gardnerLpsi}) and (\ref{gardnerMpsi}) can be written as
\begin{eqnarray}
\label{gardnerdirectcompatibility}
&& {\mathrm D}_t {\mathrm D}_x^2 \psi - {\mathrm D}_x^2 {\mathrm D}_t \psi 
= \tfrac{1}{6} 
\left(
  (6 \epsilon \pm \sqrt{-6 \beta})
  \, {\mathrm D}_x (u_t + \alpha u u_x + \beta u^2 u_x + u_{3x})
\right.
\nonumber \\
&& 
\left. 
   + ( \alpha + 2 (6 \epsilon^2 + \beta) u ) 
   ( u_t + \alpha u u_x + \beta u^2 u_x + u_{3x} ) 
   \right)
   \, \psi 
\nonumber \\
&&  
+ \, 2 \epsilon (u_t + \alpha u u_x + \beta u^2 u_x + u_{3x} ) \, {\mathrm D}_x \psi, 
\end{eqnarray}
where (\ref{gardnerLpsi}) and (\ref{gardnerMpsi}) were used repeatedly 
to eliminate $ {\mathrm D}_{xt} \psi $, $\, {\mathrm D}_t \psi $, 
$\, {\mathrm D}_x^5 \psi $, 
$\, {\mathrm D}_x^4 \psi $, $\,{\mathrm D}_x^3 \psi $, 
and $ {\mathrm D}_x^2 \psi $, in that order. 
From (\ref{gardnerdirectcompatibility}), it is clear that \eqref{gardnerLpsi} 
and \eqref{gardnerMpsi} will only be compatible on solutions of (\ref{gardner}).

The compatibility of the equations in (\ref{LaxLandM})
may be expressed directly in terms of the operators 
${\mathcal L}$ and ${\mathcal M}$ as follows
\begin{equation}
\label{derivtLpsi}
{\mathrm D}_t \left( {\mathcal L} \psi \right)
= {\mathcal L}_t \psi  + {\mathcal L} ( {\mathrm D}_t \psi ) 
= \lambda {\mathrm D}_t \psi,
\end{equation}
where
\begin{equation}
\label{Lderivtpsi}
{\mathcal L}_t \psi \equiv 
{\mathrm D}_t \left({\mathcal L} \psi \right) 
- {\mathcal L} \left( {\mathrm D}_t \psi \right).
\end{equation}
With (\ref{LaxLandM}), one has
\begin{equation}
\label{compatibilityLMpsi}
{\mathcal L}_t \psi + {\mathcal L} \left( {\mathcal M} \psi \right) 
= \lambda {\mathcal M} \psi = {\mathcal M} \left( \lambda \psi \right)
= {\mathcal M} \left( {\mathcal L} \psi \right). 
\end{equation}
For a {\em non-trivial} Lax pair for \eqref{scalarpde} to exist,
(\ref{compatibilityLMpsi}) should vanish 
{\em only on solutions} of \eqref{scalarpde}. 
If (\ref{laxeq}) is satisfied identically, i.e.,~without evaluation on 
solutions of the PDE, then the Lax operators ${\mathcal L}$ and ${\mathcal M}$ are considered~{\em trivial}). 
% MDPI: Footnote format is not allowed. we move it here. please confirm.
% WH: FOOTNOTE issue resolved. Rephrased in two sentences 
% 
Rearranging the terms yields
\begin{equation}
\label{compatibilityLMpsi2}
\left( {\mathcal L}_t + {\mathcal L} \circ {\mathcal M} 
- {\mathcal M} \circ {\mathcal L} \right) \psi 
= \left( {\mathcal L}_t 
  + \left[ {\mathcal L}, {\mathcal M} \right] \right) \psi 
  \, \dot{=} \, 0, 
\end{equation}
or, expressed in operator form by suppressing the $\psi$,
\begin{equation}
\label{laxeq}
{\mathcal L}_t + \left[ {\mathcal L}, {\mathcal M} \right] 
 \, \dot{=} \, {\mathcal O},
\end{equation}
where $\dot{=}$ means that equality holds only on solutions
of the original PDE (\ref{scalarpde}).
As before, $[ \, , \, ]$ is the commutator of the operators, and 
${\mathcal O}$ is the zero operator.
Equation~(\ref{laxeq}) is called the Lax equation.
We now show how the Lax pair $({\mathcal L},\, {\mathcal M})$ 
can be computed using the method discussed in~\cite{larue-msthesis-2011}.
\vskip 5pt
\noindent
{\bf Step 1}: Construct a candidate for 
${\mathcal L}$ and ${\mathcal M}$. 
Since the KdV and mKdV equations are special cases of (\ref{gardner}), 
it makes sense to search for ${\mathcal L}$ of rank $2$ and 
${\mathcal M}$ of rank $3$.
To construct ${\mathcal L}$, list all monomials in 
${\mathrm D}_x, u$, and~$\alpha$ of rank $2$ or less, i.e.,~
$\{ 
  \{ {\mathrm D}_x^2$, 
     $u {\mathrm D}_x$, 
     $\alpha {\mathrm D}_x$, 
     $u^2 {\mathrm I}$, 
     $\alpha u {\mathrm I}
\}$, 
 $ \{ {\mathrm D}_x$, 
    $u {\mathrm I}
 \}  
\}$, 
where the trivial terms $\alpha {\mathrm I}$ and $\alpha^2 {\mathrm I}$
have been removed. 
As explained in Section~\ref{recursion-operator}, 
apply ${\mathrm D}_x$ to the elements in the second sublist
and, as~in (\ref{Dx2onuI}),
propagate ${\mathrm D}_x$ to the right,
yielding $\{ {\mathrm D}_x^2, u_x {\mathrm I}, u {\mathrm D}_x \}$.
After removing duplicates, linearly combine the monomials from 
both sublists with constant coefficients to get a candidate for ${\mathcal L}$:
\begin{equation}
\label{gardnercandidateL}
{\mathcal L} 
= {\mathrm D}_x^2 
 + ( c_1 u + \alpha c_2 ) \, {\mathrm D}_x 
 + ( c_3 u^2 + \alpha c_4 u  + c_5 u_x ) \, {\mathrm I},
\end{equation}
where the coefficient of ${\mathrm D}_x^2$ has been set to one 
(for normalization). 

To make a candidate for ${\mathcal M}$, list all monomials in 
${\mathrm D}_x, u$, and~$\alpha$ of rank $3$ or less, i.e.,~
$\{ 
  \{ {\mathrm D}_x^3$, 
     $u {\mathrm D}_x^2$, 
     $\alpha {\mathrm D}_x^2$, 
     $u^2 {\mathrm D}_x$, 
     $\alpha u {\mathrm D}_x$, 
     $\alpha^2 {\mathrm D}_x$,  
     $u_x {\mathrm D}_x$, 
     $u^3 {\mathrm I}$,
     $\alpha u^2 {\mathrm I}$,
     $\alpha^2 u {\mathrm I} 
\}$,  
 $\{ {\mathrm D}_x^2$, 
     $u {\mathrm D}_x$, 
     $\alpha {\mathrm D}_x$, 
     $u^2 {\mathrm I}$, 
     $\alpha u {\mathrm I}
\}$, 
 $\{ {\mathrm D}_x$, 
     $u {\mathrm I}
 \}  
\}$, 
where the trivial terms $\alpha {\mathrm I}$, $\alpha^2 {\mathrm I}$, and~
$\alpha^3 {\mathrm I}$ have been removed. 
Apply ${\mathrm D}_x$ to the elements in the second sublist, yielding 
$\{  {\mathrm D}_x^3$, 
     $u_x {\mathrm D}_x$, 
     $u {\mathrm D}_x^2$, 
     $\alpha {\mathrm D}_x^2$, 
     $2 u u_x {\mathrm I}$,
     $u^2 {\mathrm D}_x$, 
     $\alpha u_x {\mathrm I}$, 
     $\alpha u {\mathrm D}_x 
\}$.
Apply ${\mathrm D}_x^2$ to the elements in the third sublist 
and use (\ref{Dx2onuI}) to obtain
$\{ {\mathrm D}_x^3$, 
    $u_{xx} {\mathrm I}$, 
    $2 u_x {\mathrm D}_x$, 
    $u {\mathrm D}_x^2
\}$.
After stripping off numerical factors and removing duplicates,
linearly combine the resulting monomials with constant coefficients
to obtain a candidate for ${\mathcal M}$:
\begin{eqnarray}
\label{gardnercandidateM}
{\mathcal M} 
&=& 
c_6 {\mathrm D}_x^3 
+ ( c_7 u + \alpha c_8 ) \, {\mathrm D}_x^2  
+ ( c_9 u^2 + \alpha c_{10} u + \alpha^2 c_{11} + c_{12} u_x ) \, {\mathrm D}_x
\nonumber \\ 
&& 
+ ( c_{13} u^3 + \alpha c_{14} u^2 + \alpha^2 c_{15} u 
+ c_{16} u u_x + \alpha c_{17} u_x + c_{18} u_{xx}) \, {\mathrm I}. 
\end{eqnarray}
\vskip 2pt
\noindent
{\bf Step 2}: Compute the undetermined coefficients.  
First, compute
\begin{eqnarray}
\label{gardnerLt}
{\mathcal L}_t 
&=& c_1 u_t {\mathrm D}_x 
    + ( 2 c_3 u u_t + \alpha c_4 u_t + c_5 u_{xt} ) \, {\mathrm I} 
\nonumber \\
&=& c_1 F {\mathrm D}_x 
    + ( 2 c_3 u F + \alpha c_4 F + c_5 {\mathrm D}_x F ) 
    \, {\mathrm I}, 
\end{eqnarray}
which, after~substituting $F$ from (\ref{gardnerF}) 
to replace $u_t$ and $u_{xt}$, yields $14$ terms.
Next, compute
\begin{eqnarray}
\label{gardnerLM}
{\mathcal L} \circ {\mathcal M} 
&=& 
 \left( {\mathrm D}_x^2 
      + ( c_1 u + \alpha c_2 ) \, {\mathrm D}_x 
      + ( c_3 u^2 + \alpha c_4 u  + c_5 u_x ) \, {\mathrm I} 
 \right) \circ
\nonumber \\
&& \left( c_6 {\mathrm D}_x^3 
       + (c_7 u + \alpha c_8) \, {\mathrm D}_x^2 
       + (c_9 u^2 + \alpha c_{10} u + \alpha^2 c_{11} 
          + c_{12} u_x) \, {\mathrm D}_x
   \right.
\nonumber \\
&& \left. 
     + ( c_{13} u^3 + \alpha c_{14} u^2 + \alpha^2 c_{15} u 
        + c_{16} u u_x + \alpha c_{17} u_x + c_{18} u_{xx} ) \, {\mathrm I}
   \right), 
\end{eqnarray}
which upon expansion has $125$ terms, 
and
\begin{eqnarray}
\label{gardnerML} 
{\mathcal M} \circ {\mathcal L} 
&=& 
 \left( 
   c_6 {\mathrm D}_x^3 
   + (c_7 u + \alpha c_8) \, {\mathrm D}_x^2  
   + (c_9 u^2 + \alpha c_{10} u + \alpha^2 c_{11} 
     + c_{12} u_x) \, {\mathrm D}_x
  \right.
\nonumber \\ 
&& \left. 
     + (c_{13} u^3 + \alpha c_{14} u^2 + \alpha^2 c_{15} u 
     + c_{16} u u_x + \alpha c_{17} u_x + c_{18} u_{xx}) \, {\mathrm I}
   \right) \circ
\nonumber \\
&&  \left( 
      {\mathrm D}_x^2 + (c_1 u + \alpha c_2) \, {\mathrm D}_x 
      + (c_3 u^2 + \alpha c_4 u  + c_5 u_x) \, {\mathrm I} 
    \right),
\end{eqnarray}
which after expansion has $126$ terms. 

Substitute (\ref{gardnerLt})--(\ref{gardnerML}) into (\ref{laxeq}). 
The resulting expression (with $105$ terms) should be identically 
equal to zero for any function $\psi(x,t)$. 
Set the coefficients of $\psi, \psi_x$, and $\psi_{xx}$ equal to zero, 
and then set the coefficients of all monomials in $u$ and its $x$-derivatives
separately equal to zero. 
This yields a {\em nonlinear} system of $24$ equations 
for the undetermined coefficients:
\begin{eqnarray}
\label{gardnerLaxpairnonlinearsystem}
&& 
2 c_7 - 3 c_1 c_6  = 0, \,\,\,
4 c_9 - 6 c_3 c_6 - c_1 c_7 = 0,
\nonumber \\
&& \phantom{c_7 - 3 c_1 c_6  = 0,} 
\vdots \phantom{4 c_9 - 6 c_3 c_6 - c_1 c_7 = 0,}
\nonumber \\
&& 
c_{12} + 2 c_{18} - c_1 - c_1 c_6 - 3 c_5 c_6 = 0, \,\,\,
\alpha (c_{17} - c_4 + c_2 c_{18} - c_4 c_6 - c_5 c_8 ) = 0.
\end{eqnarray}
For brevity, we have shown only a couple of the shortest equations 
(coming from $u_x {\mathrm D}_x^3 $ and $ u u_x {\mathrm D}_x^2,$ respectively), 
and two of the longest equations (coming from 
$ u_{3x} {\mathrm D}_x $ and 
$ u_{3x} {\mathrm I} $, respectively). 
Since each equation has a mixture of linear and 
and nonlinear terms, several solution branches occur. 
{\em Mathematica}'s \verb|Solve| function returns five non-trivial solutions. 
Three of these solutions lead to Lax pairs of lower order or degenerate Lax pairs
which will not be discussed.  
Instead, we focus on the two solutions that lead to Lax pairs that are 
useful in, e.g.,~the application of the inverse scattering transform (IST) 
and the Riemann--Hilbert methods to solve the Gardner equation. 
They have coefficients
\begin{eqnarray}
\label{gardnerLaxpairc}
&& c_2 = \frac{6 c_4 - 1}{3 c_1}, \;
c_3 = \tfrac{1}{12}(3 c_1^2 + 2 \beta), \;
c_5 = \tfrac{1}{6}(3 c_1 \pm \sqrt{-6\beta}), \;
c_6 = - 4, \;
c_7 = - 6 c_1, \;
\nonumber \\
&&
c_9 = -( 3 c_1^2 + \beta ), \;
c_{10} = c_1 c_8 - 1, \;
c_{11} = \frac{(1 - 6 c_4)(1 - 6 c_4 - c_1 c_8)}{3 c_1^2}, \;
\nonumber \\
&& 
c_{12} = - (6 c_1 \pm \sqrt{-6\beta}), \;
c_{13} = -\tfrac{1}{6} c_1 (3 c_1^2 + 2 \beta), \;
\nonumber \\
&& 
c_{14} = -\frac{3 c_1^2 + 2 \beta (1-6 c_4) -c_1 c_8 (3 c_1^2 + 2\beta)}{12 c_1},\;
\nonumber \\
&& c_{15} = \frac{c_4 (c_1 c_8 + 6 c_4 - 1)}{c_1}, \;
c_{16} = -\tfrac{1}{2} (6 c_1^2 + 2 \beta \pm c_1 \sqrt{-6\beta}), 
\;
\nonumber \\
&&
c_{17} = \frac{3 c_1 (c_1 c_8 - 1) \pm (c_1 c_8 + 6 c_4 - 1) 
         \sqrt{-6\beta} }{6 c_1}, \; 
c_{18} = - \tfrac{1}{2} (3 c_1 \pm \sqrt{-6\beta}),
\end{eqnarray}
where $c_1, c_4,$ and $c_8$ are arbitrary constants.
To be able to obtain the Lax pair for the KdV equation 
where $c_1 = 0$, 
one should require that $c_4 = \tfrac{1}{6}$ and $c_8 = 0$, 
otherwise $c_2$, $c_{11}$, $c_{14}$, and $c_{17}$ would become infinite.  
Notice that both requirements allow one to clear $c_1$ from all denominators. 
Furthermore, the~coefficients then simplify into
\begin{eqnarray}
\label{gardnerLaxpaircsimp}
&& c_2 = 0, \;\,
c_3 = \tfrac{1}{12}(3 c_1^2 + 2 \beta), \;\,
c_4 = \tfrac{1}{6}, \;\,
c_5 = \tfrac{1}{6} (3 c_1 \pm \sqrt{-6\beta}), \;\,
c_6 = -4, \;\,
c_7 = -6 c_1, \;\,
\nonumber \\
&& 
c_8 = 0, \;\,
c_9 = -(3 c_1^2 + \beta ), \;\,
c_{10} = -1, \;\,
c_{11} = 0, \;\,
c_{12} = - (6 c_1 \pm \sqrt{-6\beta}), \;\,
\nonumber \\
&& 
c_{13} = -\tfrac{1}{6} c_1 (3 c_1^2 + 2 \beta), \;\,
c_{14} = -\tfrac{1}{4} c_1, \;\,
c_{15} = 0, \;\,
c_{16} = -\tfrac{1}{2} (6 c_1^2 + 2\beta \pm c_1 \sqrt{-6\beta}), \;\,
\nonumber \\
&&
c_{17} = -\tfrac{1}{2}, \;\, 
c_{18} = - \tfrac{1}{2} (3 c_1 \pm \sqrt{-6\beta}).
\end{eqnarray}
Finally, substitute the coefficients into (\ref{gardnercandidateL}) and 
(\ref{gardnercandidateM}) to get 
% GOOD VERIFIED
\begin{eqnarray}
\label{gardnerLfinal}
{\mathcal L} &=&  
 {\mathrm D}_x^2 
 + c_1 u \, {\mathrm D}_x 
 + \tfrac{1}{6} \left( 
    \tfrac{1}{2} (3 c_1^2 + 2 \beta) u^2 
    + \alpha u  
    + (3 c_1 \pm \sqrt{-6\beta}) u_x 
 \right) \, {\mathrm I}, 
\\
\label{gardnerMfinal}
{\mathcal M} &=& 
- 4 {\mathrm D}_x^3 
- 6 c_1 u \, {\mathrm D}_x^2  
- \left(
  (3 c_1^2 + \beta ) u^2 
  + \alpha u  
  + (6 c_1 \pm \sqrt{-6\beta}) u_x 
  \right) \, {\mathrm D}_x
\nonumber \\ 
&& 
-\left( 
 \tfrac{1}{6} c_1 (3 c_1^2 + 2 \beta) u^3 
 + \tfrac{1}{4} \alpha c_1 u^2  
 + \tfrac{1}{2} (6 c_1^2 + 2 \beta \pm c_1 \sqrt{-6\beta}) u u_x 
 \right. 
\nonumber \\
&&
  \left. + \tfrac{1}{2} \alpha u_x 
  + \tfrac{1}{2} (3 c_1 \pm \sqrt{-6\beta}) u_{xx}
 \right) \, {\mathrm I},  
\end{eqnarray}
where the constant $c_1$ is arbitrary. 
Hence, this is a one-parameter family of Lax pairs. 
Set $c_1 = 2 \epsilon$ to get (\ref{gardnerLresult}) and (\ref{gardnerMresult}).
With (\ref{gardnerLfinal}) and (\ref{gardnerMfinal})
\begin{eqnarray}
\label{gardnerlaxeqeval}
&& {\mathcal L}_t + [ {\mathcal L}, {\mathcal M}] 
= \tfrac{1}{6} 
  \left(
  (3 c_1 \pm \sqrt{-6 \beta}) \, 
  {\mathrm D}_x (u_t + \alpha u u_x + \beta u^2 u_x + u_{3x})
  \right. 
\nonumber \\
&& \left. + 
   ( \alpha + (3 c_1^2 + 2 \beta) u ) 
          ( u_t + \alpha u u_x + \beta u^2 u_x + u_{3x} ) 
   \right) 
   \, {\mathrm I}
\nonumber \\
&&  + c_1 (u_t + \alpha u u_x + \beta u^2 u_x + u_{3x} ) 
\, {\mathrm D}_x,
\end{eqnarray}
which, after~setting $c_1 = 2\epsilon$, 
is equivalent to (\ref{gardnerdirectcompatibility}).

% BILINEAR FORM 
\section{Bilinear~Form}
\label{bilinear-form}
In this section we show how the Gardner equation~(\ref{gardner}) can be transformed 
into the mKdV Equation~(\ref{mkdv}) and how that helps with deriving Hirota's 
bilinear representation~\cite{hirota-book-2004} and, 
eventually, solitary wave and soliton solutions of (\ref{gardner}). 
The existence of multi-soliton solutions (i.e., soliton solutions of any order) 
is yet another proof that the Gardner equation is completely~integrable. 

A simple shift of $u$ allows one to remove the quadratic term $\alpha u u_x$.  
Indeed, set \mbox{$u = {\tilde{u}} - \frac{\alpha}{2\beta}$} to replace (\ref{gardner}) by
\begin{equation}
\label{mkdvutilde} 
 {\tilde{u}}_t - \frac{\alpha^2}{4\beta} {\tilde{u}}_x
 + \beta {\tilde{u}}^2 {\tilde{u}}_x + {\tilde{u}}_{3x} = 0,  
\end{equation}
which is still in the original independent variables $x$ and $t$.
As will be shown below, the~linear term in ${\tilde{u}}_x$ can also 
be removed by a change of independent variables. 
\vskip 5pt
\noindent
{\bf Step 1}: Construct a bilinear form as follows:
first, integrate (\ref{mkdvutilde}) with respect to $x$,
\begin{equation}
\label{mkdvint}
\partial_t \left( \int^x {\tilde{u}} \, dx \right) 
  - \frac{\alpha^2}{4\beta} {\tilde{u}}
  + \tfrac{1}{3} \beta {\tilde{u}}^3 + {\tilde{u}}_{xx} = 0, 
\end{equation}   
setting the integration ``constant'' $c(t)$ equal to zero. 
As with the mKdV equation~\cite{hereman-goktas-springer-2024}, 
substitute the Hirota transformation 
% 
% (the transformation is motivated by 
% (\ref{gardnertruncatedlaurentseries})
% but, to derive a bilinear form, $g$ should be replaced by 
% $\frac{f-ig}{f+ig}$ as explained in~\cite{hereman-goktas-springer-2024})
% MDPI: Footnote format is not allowed. we move it here. please confirm.
% WH: FOOTNOTE issue resolved. To keep the flow of the text, we have 
% removed the text of the original footnote. 
\begin{equation}
\label{mkdvtffinal}
{\tilde{u}}  
 = i \sqrt{\frac{6}{\beta}} \Biggl( \ln \left( \frac{f+ig}{f-ig} \right) \Biggr)_x
 = 2 \sqrt{\frac{6}{\beta}} \Biggl( {\mathrm{Arctan}} \left( \tfrac{f}{g} \right) \Biggr)_x
 = 2 \sqrt{\frac{6}{\beta}} \Biggl( \frac{f_x g - f g_x}{f^2 + g^2} \Biggr)
\end{equation}
into (\ref{mkdvint}). Then, divide by $-2\sqrt{\tfrac{6}{\beta}}$ 
and regroup the terms, yielding
\begin{eqnarray}
\label{gardnerinfg}
&& \left( f^2 + g^2 \right)
   \left( 
   g f_t - f g_t  + 3 f_x g_{xx} - 3 g_x f_{xx} - f g_{3x} + g f_{3x} 
   - \frac{\alpha^2}{4\beta} \left(f g_x - g f_x \right)
   \right) 
\nonumber \\
&& - 6 \left( g f_x - f g_x) (f f_{xx} + g g_{xx} - f_x^2 - g_x^2
    \right) = 0. 
\end{eqnarray}
Next, set the factors multiplying $f^2+g^2$ and $g f_x-f g_x$ separately 
equal to zero, to~obtain
\begin{eqnarray}
\label{gardnerinfgfirst}
&& g f_t - f g_t  + 3 f_x g_{xx} - 3 g_x f_{xx} - f g_{3x} + g f_{3x} 
   - \frac{\alpha^2}{4\beta} \left(f g_x - g f_x \right) = 0,
\\
\label{gardnerinfgsecond}
&& f f_{xx} + g g_{xx} - f_x^2 - g_x^2 = 0.
\end{eqnarray}
Based on the scaling symmetry of (\ref{gardner}) with weights 
(\ref{gardnerweights}) and the structure of the bilinear form 
of the mKdV equation~\cite{hereman-goktas-springer-2024}, 
recast the above equations in bilinear form,
\begin{eqnarray}
\label{gardnerbilinearfirst}
&& (c_1 D_t + c_2 D_x^3 + \alpha^2 c_3 D_x) (f \mathbf{\cdot} g) = 0,
\\
\label{gardnerbilinearsecond}
&& c_4 D_x^2 (f \mathbf{\cdot} f + g \mathbf{\cdot} g) = 0, 
\end{eqnarray}
with undetermined coefficients $c_1, c_2, c_3$, and $c_4$. 
Notice that with $W(f) = W(g) = 0$, all terms in (\ref{gardnerinfgfirst}) and (\ref{gardnerbilinearfirst}) have weights of three, whereas the terms in (\ref{gardnerinfgsecond}) and (\ref{gardnerbilinearsecond}) 
have weights of~two. 

The bilinear operators $D_x$ and $D_t$ (not to be confused with total 
derivatives used in earlier sections) are defined as
\begin{eqnarray}
\label{hirotaDx}
D_x^m (f \mathbf{\cdot} g )
\!\!&=&\!\! {( \partial_{x} - \partial_{x'} )}^m  f(x,t) g(x',t) \bigg|_{x'=x}
 = \sum_{j=0}^m \frac{(-1)^{m-j}m!} {j!(m - j)!}
  \left(\frac{\partial^j f} {\partial {x^j}} \right)
  \left(\frac{\partial^{m-j} g} {\partial {x^{m-j}}} \right),
\\
D_t^n (f \mathbf{\cdot} g )
\!\!&=&\!\! {(\partial_{t} - \partial_{t'} )}^n f(x,t) g(x,t') \bigg|_{t'=t}
 = \sum_{j=0}^n \frac{(-1)^{n-j}n!} {j!(n - j)!}
  \left(\frac{\partial^j f} {\partial {t^j}} \right)
  \left(\frac{\partial^{n-j} g} {\partial {t^{n-j}}} \right),  
\end{eqnarray}
which represent the Leibniz rule for $x$-derivatives (and $t$-derivatives, respectively) 
of products of functions with every other sign flipped. 
Explicitly,
% WH Oct 2, 2024 
% WH: Retracted some space between equation and equation number 
\begin{eqnarray}
\label{gardnerbilinearfirstexplicit}
(c_1 D_t + c_2 D_x^3 + \alpha^2 c_3 D_x) (f \mathbf{\cdot} g) 
\!\!&=&\!\! c_1 (g f_t - f g_t) 
  + c_2 (3 f_x g_{xx} - 3 g_x f_{xx} + g f_{3x} - f g_{3x})
\nonumber \\
\!\!&&\!\! + \alpha^2 c_3 (g f_x - f g_x),
\\
\label{gardnerbilinearsecondexplicit}
c_4 D_x^2 (f \mathbf{\cdot} f + g \mathbf{\cdot} g) 
\!\!&=&\!\!2 c_4 (f f_{xx} + g g_{xx} - f_x^2 - g_x^2).
\end{eqnarray}
\vskip 5pt
\noindent
{\bf Step 2}: Compute the undetermined coefficients. 
Equate (\ref{gardnerbilinearfirst}) and (\ref{gardnerbilinearfirstexplicit}) 
and treat all monomials in $f$ and $g$ and their derivatives as independent 
to get $c_1 = c_2 = 1,$ and $c_3 = \frac{1}{4\beta}.$
Perform the same with (\ref{gardnerbilinearsecond}) and (\ref{gardnerbilinearsecondexplicit}) 
to get $c_4 = \frac{1}{2}$. 
Finally, substitute the constants into (\ref{gardnerbilinearfirst}) 
and (\ref{gardnerbilinearsecond}) and clear common factors to~get a
bilinear representation of (\ref{mkdvutilde}):
\begin{equation}
\label{gardnerbilinearfinal}
(D_t + D_x^3 + \frac{\alpha^2}{4 \beta} D_x) (f \mathbf{\cdot} g) 
= 0, 
\quad
D_x^2 (f \mathbf{\cdot} f + g \mathbf{\cdot} g) = 0,
\end{equation}
which, in~light of (\ref{mkdvtffinal}), will only lead to real solutions for the focusing Gardner equation $(\beta > 0)$. 

The above bilinear formulation is expressed in the original variables $x$ and $t$.
Of course, the~term $\frac{\alpha^2}{4 \beta} D_x$ in (\ref{gardnerbilinearfinal}) 
can be removed at the cost of introducing a new variable $X$.
Indeed, using the chain rule for differentiation, 
one can readily verify that the Galilean transformation
$(x, t, u(x,t)) \longrightarrow (X, T, U(X,T))$ where
\begin{equation}
\label{gardnertfmkdv}
X = x + \frac{\alpha^2}{4\beta} \, t, 
\quad
T = t, 
\quad
u(x,t) = U(X,T) - \frac{\alpha}{2\beta},  
\end{equation}
takes (\ref{gardner}) into
\begin{equation}
\label{mkdvXT} 
U_T + \beta \, U^2 U_X + U_{3X} = 0,  
\end{equation}
with bilinear representation~\cite{hereman-goktas-springer-2024}
\begin{equation}
\label{gardnerbilinearmkdv}
(D_T + D_X^3) (F \mathbf{\cdot} G) = 0, \quad
D_X^2 (F \mathbf{\cdot} F + G \mathbf{\cdot} G) = 0.
\end{equation}
Once particular solutions $F(X,T)$ and $G(X,T)$ of
\begin{eqnarray}
\label{gardnerinFGfirst}
&& G F_t - F G_t  + 3 F_x G_{xx} - 3 G_x F_{xx} - F G_{3x} + G F_{3x} = 0,
\\
\label{gardnerinFGsecond}
&& F F_{xx} + G G_{xx} - F_x^2 - G_x^2 = 0,
\end{eqnarray}
are computed,
\begin{equation}
\label{mkdvtfinFG}
U(X,T) = 2 \sqrt{\frac{6}{\beta}} \left( \frac{F_X G - F G_X}{F^2 + G^2} \right)
\end{equation}
will solve (\ref{mkdvXT}). 
Solutions $u(x,t)$ of (\ref{gardner}) then follow from
\begin{equation}
\label{gardnersolu}
u(x,t) = 2 \sqrt{\frac{6}{\beta}} \left( \frac{F_X G - F G_X}{F^2 + G^2} \right) 
- \frac{\alpha}{2\beta}, 
\end{equation}
after using (\ref{gardnertfmkdv}) to return to the original variables 
$u$, $x$, and~$t$.
\section{Gardner~Transformation}
\label{gardner-transformation}
In this section we discuss a slight generalization of the Miura transformation
~\cite{miura-i-jmp-1968}, 
sometimes called the Gardner transformation or Gardner 
% WH Oct 2, 2024 corrected gardner-etal-ii-jmp-1968 --> miura-etal-ii-jmp-1968
% and order adjusted 
transform~\cite{alejo-etal-tams-2012,miura-siamrev-1976,miura-etal-ii-jmp-1968,munoz-dcds-2016},
\begin{equation}
\label{gardnertfgardnertokdv}
u = \frac{\beta}{\gamma} 
    \left (
     v^2 \pm \sqrt{-\tfrac{6}{\beta}} \, v_x + \frac{\alpha}{\beta} v
    \right), 
\end{equation}
which connects solutions $v(x,t)$ of the Gardner equation,
\begin{equation}
\label{gardnerinv} 
v_t + \alpha v v_x + \beta v^2 v_x + v_{3x} = 0,
\end{equation}
with $\beta \ne 0$, to~solutions $u(x,t)$ of the KdV equation,
\begin{equation}
\label{kdvwithgamma} 
u_t + \gamma u u_x + u_{3x} = 0,
\end{equation}
with arbitrary coefficient $\gamma \ne 0$. 
The use of $\gamma$ will avoid confusion with $\alpha$ in (\ref{gardnerinv}) 
and, more importantly, allow us to set $\alpha = 0$ to get the standard Miura transformation (see Section~\ref{mkdv-equation}). 

Substituting (\ref{gardnertfgardnertokdv}) into (\ref{kdvwithgamma}), 
it is straightforward to verify that
\begin{equation}
\label{gardnerconnectkdvgardner}
u_t + \gamma u u_x + u_{3x} 
\,\, \dot{=} \,\, \tfrac{\beta}{\gamma} 
 \left(
  ( 2 v + \tfrac{\alpha}{\beta} ) \, {\mathrm I} 
  \, \pm \sqrt{-\tfrac{6}{\beta}} \, {\mathrm D}_x 
 \right)
 \left( v_t + \alpha v v_x + \beta v^2 v_x + v_{3x} \right), 
\end{equation} 
where $\,\dot{=}\,$ means that the left hand side is evaluated on solutions 
of (\ref{gardnerinv}). 
As before, ${\mathrm I}$ is the identity operator and ${\mathrm D}_x$ is 
the total derivative operator defined in (\ref{gardnerDxJ}).
Clearly, the~Gardner transformation will only be real for the defocusing 
Gardner~equation.

We will show how (\ref{gardnertfgardnertokdv}) can be computed using 
the scaling symmetries of the KdV and Gardner equations 
discussed in Section~\ref{scaling-symmetry}.
Recall that $W(u) = 2$, $W(v) = W(\alpha) = 1$, and~$W(\gamma) = W(\beta) = 0$.
Therefore, (\ref{gardnertfgardnertokdv}) is uniform in rank 
of rank two. 
\vskip 5pt
\noindent
{\bf Step 1}: Construct a candidate for the Gardner transformation. 
Make the list $\{ \{v^2, \alpha v, \alpha^2 \}$, $\{v, \alpha \} \}$ 
of monomials in $v$ and $\alpha$ of rank $2$ or less.
By differentiating its elements, replace the second sublist by $\{ v_x \}$. 
Linearly combine the resulting elements with undetermined coefficients,
\begin{equation}
\label{gardnertfcandidate}
u = c_1 v^2 + \alpha c_2 v + \alpha^2 c_3 + c_4 v_x, 
\end{equation}
to generate a candidate for the Gardner transformation. 
\vskip 5pt
\noindent
{\bf Step 2}: Compute the undetermined coefficients. 
Substitute (\ref{gardnertfcandidate}) into (\ref{kdvwithgamma}) and, 
using (\ref{gardnerinv}), replace $v_t$ and $v_{tx}$. 
The resulting expression must vanish on the jet space, leading to 
$c_2 c_3 = c_3 c_4 = 0$ and half a dozen more complicated equations. 
For $c_4 = 0$, (\ref{gardnertfcandidate}) would become an algebraic 
transformation. 
Hence, $c_3 = 0$ and, after~simplification of the more complicated equations, 
one is left with the following {\em nonlinear} system
\begin{eqnarray}
\label{gardnertfsystem}
\gamma c_1 - \beta = 0, \;\, 
\gamma c_2 - 1 = 0, \;\, 
6 c_1 + \gamma c_4^2 = 0, \;\,
3 \gamma c_1 c_2 - \beta c_2 - 2 c_1 = 0.
\end{eqnarray}
Substitute the solution,  
$c_1 = \tfrac{\beta}{\gamma}$, 
$\, c_2 = \tfrac{1}{\gamma}$, 
$\, c_4 = \pm \frac{1}{\gamma} \sqrt{-6\beta}$,
and $c_3 = 0$, into~(\ref{gardnertfcandidate}) 
to get (\ref{gardnertfgardnertokdv}).

Of course, the~ ``uniformity-in-rank'' argument also applies to 
(\ref{gardnerconnectkdvgardner}), and therefore can be used to derive that equation.
Observe that the ranks of the left and right hand sides of 
(\ref{gardnerconnectkdvgardner}) match.
Since the terms of the KdV equation (left) and Gardner equation (right) have ranks five and four, 
respectively, the~operator that connects them must have rank one. 
Thus, only $v {\mathrm I}$, $\alpha {\mathrm I}$, and~${\mathrm D}_x$ 
can occur in that operator. 
Apply the candidate for the operator,
\begin{equation}
\label{candidategardneroper}
(C_1 v + \alpha C_2) \, {\mathrm I} + C_3 \, {\mathrm D}_x,  
\end{equation}
to (\ref{gardnerinv}), and equate the resulting expression 
to (\ref{kdvwithgamma}) after substitution of (\ref{gardnertfgardnertokdv}) 
but without evaluation on solutions of (\ref{gardnerinv}).
After simplification, this yields the {\em linear} system
\begin{eqnarray}
\label{gardneroperatorsystem}
\gamma C_1 - 2 \beta = 0, \;\, 
\gamma C_2 - 1 = 0, \;\, 
\gamma C_3 \pm \sqrt{-6\beta} = 0, \;\,
\gamma C_1 + \beta \gamma C_2 - 3 \beta = 0. 
\end{eqnarray}  
Solve the system to obtain
$ C_1 = \tfrac{2\beta}{\gamma}$, 
$ \, C_2 = \tfrac{1}{\gamma}$, 
and 
$ \, C_3 = \pm \tfrac{1}{\gamma} \sqrt{-6 \beta}$.
Substitute the solution into (\ref{candidategardneroper}) to get (\ref{gardnerconnectkdvgardner}).

Using (\ref{gardnertfgardnertokdv}), some solutions for the KdV equation 
could be obtained from those of the Gardner equation 
(see, e.g.,~\cite{kirane-etal-csf-2024}).
For $\alpha = 0$, (\ref{gardnertfgardnertokdv}) reduces to the Miura transformation, 
allowing one to generate solutions of the KdV equation from those of the mKdV equation.
Although this is worthy of investigation, it is beyond the scope of this article. 
For an in-depth discussion of connections between the KdV, mKdV, and~Gardner equations 
and additional applications of the Gardner and Miura transformations, we refer to~\cite{alejo-etal-tams-2012,kirane-etal-csf-2024,munoz-dcds-2016}.

% KDV EQUATION 
\section{The Korteweg--de Vries~Equation}
\label{kdv-equation}
Since the KdV equation is a special case of the Gardner equation,
its conservation laws, higher-order symmetries, and~recursion operator  
immediately follow from those of (\ref{gardner}) by setting 
$\beta = 0$ 
in (\ref{gardnerconslaw1})--(\ref{gardnerconslaw3}),  
(\ref{gardnersym1})--(\ref{gardnersym3}), and~(\ref{gardnerR}), respectively. 
The conservation laws for the KdV equation have been known since the 1970s~\cite{kruskal-etal-v-jmp-1970,miura-etal-ii-jmp-1968} 
and have played an important role in the development of the concept of 
{\em complete integrability}.
Its symmetries and recursion operator have been studied in, e.g.,~
\cite{olver-jmp-1977}.

The well-known Lax pair~\cite{lax-cpam-1968} for the KdV equation,
% \begin{eqnarray}
\begin{equation}
\label{kdvLMresult}
 {\mathcal L} =
   {\mathrm D}_x^2 + \tfrac{1}{6} \alpha u \, {\mathrm I}, 
   \quad
 {\mathcal M} =
      - 4 {\mathrm D}_x^3 - \alpha u  \, {\mathrm D}_x
     - \tfrac{1}{2} \alpha u_x \, {\mathrm I}, 
\end{equation}
follows from (\ref{gardnerLresult}) and (\ref{gardnerMresult}) by setting 
$\beta = \epsilon = 0$. 

The bilinear form for the KdV equation is much simpler than the one for the 
mKdV equation, and so are its soliton solutions. 
The interested reader is referred to~\cite{ablowitz-clarkson-book-1991,drazin-johnson-book-1989,hereman-goktas-springer-2024,hirota-book-2004} 
for the bilinear formulation and explicit formulas for the two-, three-, and~
$N$-soliton~solutions. 

\begin{figure}[H]

\includegraphics[width=3.65in, height=1.96in]{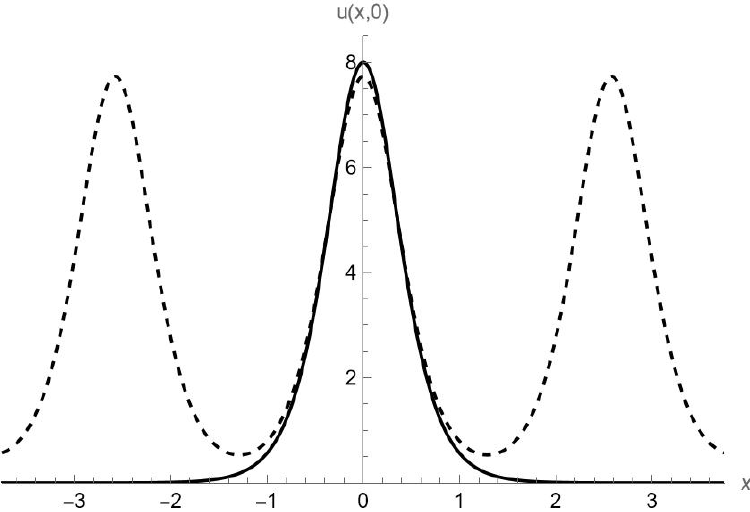}

\caption{Graphs of the solitary wave (solid line) and 
cnoidal wave (dashed line) solutions for 
$\alpha = 6, k = 2,\, m =\tfrac{9}{10},$ and $\delta = 0$.}
\label{kdv-sech-cn-sols}
\end{figure}

For completeness, we only include the solitary wave and cnoidal wave solutions, 
which were first derived in~\cite{korteweg-devries-philmag-1895},
\begin{eqnarray}
\label{kdvsolitarysimpler}
u(x,t) &=& \frac{12 k^2}{\alpha} \, {\mathrm{sech}}^2 (k x - 4 k^3 t + \delta)
    = \frac{12 k^2}{\alpha} \, 
     \left( 
     1 - \tanh^2 (k x - 4 k^3 t + \delta) 
     \right),
\\
\label{kdvcnsimpler}
u(x,t) &=& 
  \frac{8 k^2}{\alpha} (1-m) 
  + \frac{12 k^2}{\alpha} m \, {\mathrm{cn}}^2 (k x - 4 k^3 t + \delta ; m),
\end{eqnarray}
where $m \in (0,1)$ is the modulus of the Jacobi elliptic cosine 
(${\mathrm{cn}}$) function.
Both solutions are depicted in Figure~\ref{kdv-sech-cn-sols}. 
When $m$ approaches 1, the~peaks of the ${\mathrm{cn}}$-squared solution become a 
bit taller, and the~valleys become lower and flatter before they spread out 
horizontally to become the ${\mathrm{sech}}$-squared solution.
Both solutions (\ref{kdvsolitarysimpler}) and (\ref{kdvcnsimpler}) satisfy 
$\lim_{|x| \to \infty} u(x,t) = 0$ for all $t$. 
The more general expressions corresponding to a nonzero boundary condition 
can be found in, e.g.,~refs.
\cite{ablowitz-clarkson-book-1991,ablowitz-segur-book-1981,hereman-encyclopia-2008}.

The soliton solutions of (\ref{kdv}) can be computed with our code 
\verb|PDESolitonSolutions.m| 
\cite{goktas-hereman-code-solitons-2023}.
A discussion of their properties is outside the scope of this paper.
Instead, we refer to
~\cite{ablowitz-clarkson-book-1991,ablowitz-segur-book-1981,drazin-johnson-book-1989,hereman-goktas-springer-2024} and the references given in~\cite{hereman-encyclopia-2008}. 

\unskip
% 
% MKDV EQUATION 
\section{The Modified Korteweg--de Vries~Equation}
\label{mkdv-equation}
Without extra work, we have the first three conservation laws~\cite{miura-etal-ii-jmp-1968} and higher-order symmetries~\cite{olver-jmp-1977} for the mKdV equation by setting $\alpha=0$ in 
(\ref{gardnerconslaw1})--(\ref{gardnerconslaw3}) and 
(\ref{gardnersym1})--(\ref{gardnersym3}), respectively.
The recursion operator~\cite{olver-jmp-1977} connecting those symmetries follows 
from (\ref{gardnerR}) with $\alpha = 0$.

Likewise, a~one parameter Lax pair for the mKdV equation follows from 
(\ref{gardnerLresult}) and (\ref{gardnerMresult}) by setting $\alpha = 0$:
\begin{eqnarray}
\label{mkdvgardnerLresult}
{\mathcal L} \!&\!=\!&\!  
 {\mathrm D}_x^2 
 + 2 \epsilon u \, {\mathrm D}_x 
 + \tfrac{1}{6} \left( 
    (6 \epsilon^2 + \beta) u^2  
    + (6 \epsilon \pm \sqrt{-6\beta}) u_x 
 \right) \, {\mathrm I}, 
\\
\label{mkdvgardnerMresult}
{\mathcal M} \!&\!=\!&\! 
  - 4 {\mathrm D}_x^3 
  - 12 \epsilon u \, {\mathrm D}_x^2  
  - \left(
    (12 \epsilon^2 + \beta ) u^2 
    + (12 \epsilon \pm \sqrt{-6\beta}) u_x 
  \right) \, {\mathrm D}_x
\nonumber \\ 
\!&\!&\! 
 -\left( 
  \tfrac{1}{3} \epsilon (12 \epsilon^2 + 2 \beta) u^3 
  + (12 \epsilon^2 + \beta \pm \epsilon \sqrt{-6\beta}) u u_x 
  + \tfrac{1}{2} (6 \epsilon \pm \sqrt{-6\beta}) u_{xx} 
  \right) {\mathrm I},
\end{eqnarray}
which for $\epsilon = 0$ simplify into
\begin{eqnarray}
\label{mkdvgardnerLresultspecial}
{\mathcal L} &=&  
 {\mathrm D}_x^2 
 + \tfrac{1}{6} 
  \left( 
     \beta u^2 \pm \sqrt{-6\beta} u_x 
  \right) \, {\mathrm I}, 
\\
\label{mkdvgardnerMresultspecial}
{\mathcal M} &=& 
  - 4 {\mathrm D}_x^3 - (\beta  u^2 \pm \sqrt{-6\beta} u_x) \, {\mathrm D}_x
  - (\beta u u_x \pm \tfrac{1}{2} \sqrt{-6\beta} u_{xx} ) \, 
  {\mathrm I}. 
\end{eqnarray}
To our knowledge, this Lax pair first appeared in~\cite{wadati-jpsjpn-1972,wadati-jpsjpn-1973}.
A further discussion of other Lax pairs for the mKdV equation can be found in~\cite{burde-axioms-2024,hickman-etal-aa-2012}.
In the latter paper, a more comprehensive algorithm to compute Lax pairs in operator 
form is presented, with the mKdV equation among its~examples. 

% WH Oct 2, 2024 correct but the order had to be adjusted 
The Miura transformation~\cite{miura-i-jmp-1968,miura-siamrev-1976,miura-etal-ii-jmp-1968},
\begin{equation}
\label{miuratfmkdvtokdv}
u = \frac{\beta}{\gamma} 
    \left (
     v^2 \pm \sqrt{-\tfrac{6}{\beta}}\, v_x 
    \right), 
\end{equation}
which connects solutions of the mKdV equation~(\ref{mkdv}) 
(with dependent variable $v(x,t)$)
to solutions $u(x,t)$ of the KdV equation~(\ref{kdvwithgamma}) readily follows from (\ref{gardnertfgardnertokdv}) by setting 
$\alpha = 0$. 
Likewise, (\ref{gardnerconnectkdvgardner}) reduces to
\begin{equation}
\label{miuraconnectkdvmkdv}
u_t + \gamma u u_x + u_{3x} 
\; \dot{=} \; \tfrac{\beta}{\gamma} 
 \left(
  2 v \, {\mathrm I} \, \pm \sqrt{-\tfrac{6}{\beta}} \, {\mathrm D}_x 
 \right)
 \left( v_t + \beta v^2 v_x + v_{3x} \right).
\end{equation} 
Various types of solutions for both the focusing and defocusing mKdV equations have 
been reported in the literature~\cite{slyunyaev-pelinovsky-prl-2016,zhang-etal-revmathphys-2014,zhang-yan-physicad-2020}, 
including bright and dark solitons, breathers, rational solutions, kinks, etc.  
In this paper we mainly focus on 
solitary wave and soliton solutions of the mKdV equation. 
The nature of the solutions of (\ref{mkdvXT}) depends on the sign of $\beta$.
Two cases have to be considered.
\vskip 4pt
\noindent
{\bf Case I}:
% MDPI: Please check if the bold and no indent format should be retained. 
% same as below.
% WH: retain bold and no indent
The focusing mKdV equation ($\beta > 0$).
\vskip 2pt
\noindent
The mKdV Equation~(\ref{mkdvXT}) has solutions involving a $\cosh$-function,
\begin{equation}
\label{mkdvcoshsolU0nonzero}
U(X,T) = U_0 + \frac{3 k^2}{ 
      \beta \left(
      U_0 \pm \sqrt{U_0^2 + \frac{3 k^2}{2 \beta}}\, 
      \cosh \Theta 
      \right) } , 
\end{equation}
with $\Theta = k X - (\beta U_0^2 + k^2) k T + \delta$, 
where the boundary value $U_0$, wave number $k$, and~phase $\delta$ are arbitrary constants. 
To prevent blow-up in finite time we will only consider the plus sign in 
(\ref{mkdvcoshsolU0nonzero}).
% (to prevent blow-up in finite time we will only consider the plus sign)
% MDPI: Footnote format is not allowed. we move it here. please confirm.
% WH: FOOTNOTE issue resolved. We have rephrased and put the footnote 
% as a separate sentence in the text (see higher). 
The solution satisfying $\lim_{|X| \to \infty} U(X,T) = U_0$ 
has been computed with Hirota's method in~\cite{ablowitz-satsuma-jmp-1978}, 
where only the case $\beta > 0$ was considered.
Solution (\ref{mkdvcoshsolU0nonzero}) is also valid for $\beta < 0$ with a caveat (see below). 
For $U_0 = 0$ the solution reduces to the well-known ${\mathrm{sech}}$-solution,
\begin{equation}
\label{mkdvcoshsolU0zero}
U(X,T) = \pm \sqrt{\tfrac{6}{\beta}}\, k \, {\mathrm{sech}} 
          \left( k X - k^3 T + \delta \right), 
\end{equation}
where the $\pm$ sign is due to the invariance of (\ref{mkdvXT}) 
under the discrete symmetry $U \rightarrow -U$.

The soliton solutions of the focusing mKdV equation have been known for a long time and can be computed with a variety of methods. 
Adhering to Hirota's method~\cite{hereman-goktas-springer-2024,hirota-book-2004}, 
the one-soliton solution readily follows from substitution of
\begin{equation}
\label{mkdv1solFG}
F = {\mathrm{e}}^{\Theta} 
  = {\mathrm{e}}^{k X - \omega T + \delta} \quad {\mathrm{and}} \quad G = 1
\end{equation}
into (\ref{gardnerinFGfirst}), yielding
$\omega = k^3$ and (\ref{gardnerinFGsecond}), which is identically satisfied.
From (\ref{mkdvtfinFG}), one then obtains
\begin{eqnarray}
U(X,T) &=& 2 \sqrt{\frac{6}{\beta}} \left( \frac{F_x}{1+F^2} \right)
= 2\sqrt{\frac{6}{\beta}} \, 
  k \, \left( \frac{{\mathrm{e}}^{\Theta}}{1 + {\mathrm{e}}^{2\Theta}} \right)
= \sqrt{\tfrac{6}{\beta}} \, k \, {\mathrm{sech}}\, \Theta
\nonumber \\
&=& \sqrt{\tfrac{6}{\beta}} \, k \, {\mathrm{sech}} \, (k X - k^3 T + \delta)
 = 2 \sqrt{\tfrac{6}{\beta}} \, K \, {\mathrm{sech}} \left( 2 K X - 8 K^3 T + \delta \right), 
\end{eqnarray}
where $K = \frac{k}{2}$, which matches (\ref{mkdvcoshsolU0zero}). 
The two-soliton solution~\cite{hereman-goktas-springer-2024} 
follows from
\begin{equation}
\label{mkdv2solFG}
F = {\mathrm{e}}^{\Theta_1} + {\mathrm{e}}^{\Theta_2}
    \quad {\mathrm{and}} \quad 
    G = 1 - a_{12} {\mathrm{e}}^{\Theta_1 + \Theta_2},
\end{equation}
 with $\Theta_i = k_i X - k_i^3 T + \delta_i$ and 
$a_{12} = \left( \frac{k_1-k_2}{k_1+k_2} \right)^2$.
Then, from~(\ref{mkdvtfinFG})
\begin{equation}
\label{mkdvfocustwosol}
U(X,T) = 2 \sqrt{\frac{6}{\beta}}
\left(
\frac{k_1 {\mathrm{e}}^{\Theta_1} + k_2 {\mathrm{e}}^{\Theta_2}
  + a_{12} \,
    (k_1 {\mathrm{e}}^{\Theta_2} + k_2 {\mathrm{e}}^{\Theta_1})
        {\mathrm{e}}^{\Theta_1 + \Theta_2}}{
    1 + {\mathrm{e}}^{2\Theta_1} + {\mathrm{e}}^{2\Theta_2}
  + \displaystyle{\frac{8 k_1 k_2}{(k_1 + k_2)^2}}\,  
  {\mathrm{e}}^{\Theta_1 + \Theta_2}
  + a_{12}^2 \,
     {\mathrm{e}}^{2\Theta_1 + 2\Theta_2}} \right).
\end{equation}
% WH: Oct 2, 2024 I swapped the references 
Details of the derivation are given in~\cite{hereman-goktas-springer-2024}
and~\cite{hirota-book-2004}, where formulas for the three- and $N$-soliton solutions can also be found. 
\vskip 4pt
\noindent
{\bf Case II}: The defocusing mKdV equation ($\beta < 0$).
% MDPI: have repeat paragraph below. please check if it is correct.
% WH: Rephrase below to avoid repetition.
\vskip 2pt
\noindent
Using the Zakharov--Shabat method, the defocusing mKdV equation was solved in~\cite{ono-jphyssocjpn-1976}. 
Solution (\ref{mkdvcoshsolU0nonzero}) with $X$ replaced by $-X$ corresponds to 
the case $N = 1$ in~\cite{ono-jphyssocjpn-1976}.
Notice that when $\beta < 0$, solution (\ref{mkdvcoshsolU0nonzero}) 
will only exist if $k^2 < \tfrac{2 |\beta| }{3} U_0^2.$ 
This will be discussed in greater detail in the next~section. 

A new exact two-soliton solution of (\ref{mkdvXT}) with $\beta < 0$ is 
presented in~\cite{mucalica-pelinovsky-lmp-2024}:
\begin{equation}   
\label{mkdvdefoctwosol}
u(X,T) \!=\! \sqrt{-\frac{6}{\beta}}
\left(\! \frac{
 \sinh(\Phi \!+\! 2\delta){\mathrm{e}}^{-2\Psi}
 \!+\! \sinh(\Phi \!-\! 2\delta){\mathrm{e}}^{2\Psi}
 \!+\! 2 \sinh(\Phi)(1 \!-\! \sinh^2(2\delta)){\mathrm{sech}}(2\delta)
 }{
 \cosh(\Phi \!+\! 2\delta){\mathrm{e}}^{-2\Psi}
 \!+\! \cosh(\Phi \!-\! 2\delta){\mathrm{e}}^{2\Psi}
 \!+\! 2 \cosh(\Phi)\cosh (2\delta)
 } \!\right)
\end{equation}
with 
$\Phi = X + 2 T,$ 
and 
$\Psi = 
\left( X + 2 ( 1 + 2 {\mathrm{sech}}^2(2\delta) ) T + X_0 \right) \tanh(2\delta),$
with $\delta > 0$ and $X_0$ arbitrary real parameters.
This solution is obtained as the limit for the modulus going to one of 
a dark breather solution (involving Jacobi elliptic functions and elliptic integrals) 
of the defocusing mKdV~equation. 

Solutions (\ref{mkdvcoshsolU0nonzero}) and (\ref{mkdvcoshsolU0zero}) will be used 
in the next section to find table-top and hump-shaped solutions of (\ref{gardner}). 
In turn, solution (\ref{mkdvdefoctwosol}) will lead to a two-soliton solution of the defocusing Gardner~equation. 

% SOLITARY WAVES
\section{Solitary Wave and Periodic~Solutions}
\label{gardner-solitary-wave-solutions}
As pointed out in~\cite{kamchatnov-etal-pre-2012}, the~fact that (\ref{gardner}) 
is invariant under the transformation
\begin{equation}
\label{gardnerinvariance}
u \rightarrow -\left( u + \tfrac{\alpha}{\beta} \right)
\end{equation}
makes it possible to have solutions of different polarity for that equation, 
most notably, ``bright'' as well as ``dark'' solitons, depending on the initial conditions. 
However, a~solution that vanishes at $x = \pm \infty$ will be transformed by (\ref{gardnerinvariance}) into one that goes 
to $-\tfrac{\alpha}{\beta}$ as $x \rightarrow \pm \infty$.
In this section we only cover a subset of the many types of solutions that (\ref{gardner}) admits~\cite{pelinovsky-etal-fluids-2022}.

\begin{figure}[H]

\begin{tabular}{ccc}
\includegraphics[width=2.1875in,height=1.9625in]{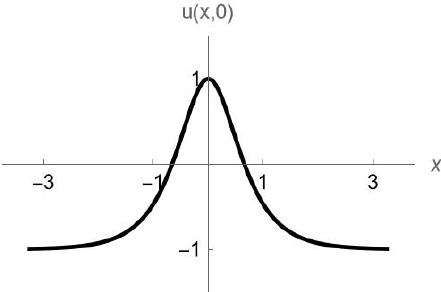}
& \phantom{x} & 
\includegraphics[width=2.1875in,height=1.9625in]{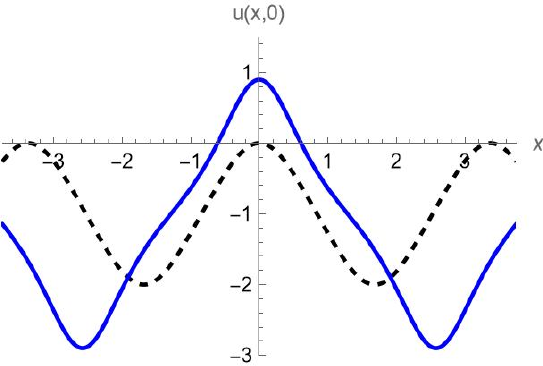}
\end{tabular}
\caption{Graphs of (\ref{sechsolsfocgardner}) (\textbf{left}) and (\ref{cnsolfocgardner}) (\textbf{right}) both with the minus signs in front 
of the square roots, and~both for $\alpha = 12$, $\beta = 6$, 
$\delta = 0$, and~$k = 2$.  The curves on the right correspond to 
$m = 0.25$ (dashed line) and $m = 0.9$ (solid line). }
\label{graph-gardnerfocussechcnsols}
\end{figure}

\begin{figure}[H]

\begin{tabular}{ccc}
\includegraphics[width=2.1875in,height=1.9625in]{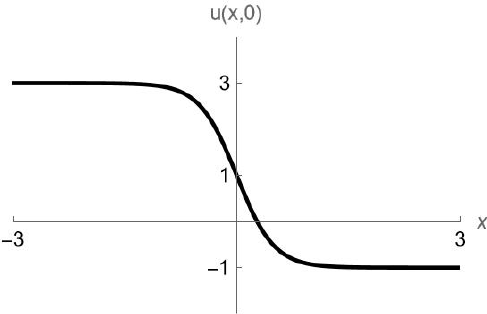}
& \phantom{x} & 
\includegraphics[width=2.1875in,height=1.9625in]{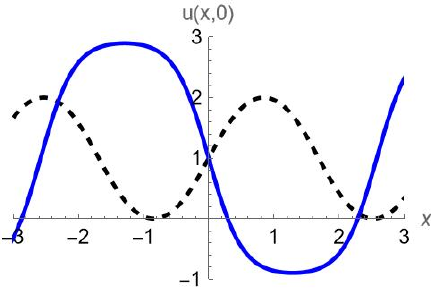}
\end{tabular}
% 
% MDPI: Some content is overlap. Please check if it affect reading.
% WH: All good the way it is. 
\caption{Graphs of (\ref{tanhsolsdefocgardner}) (\textbf{left}) and (\ref{snsolsdefocgardner}) (\textbf{right}), 
both for $\alpha = 12$, $\beta = -6$, $\delta = 0$, and~$k = 2$. 
The curves on the right correspond to $m = 0.25$ (dashed line) and 
$m = 0.9$ (solid line). }
\label{graph-gardnerdefocustanhsnsols}
\end{figure}

Depending on the sign of $\beta$ in (\ref{gardner}), it is straightforward~\cite{wazwaz-cnsns-2007} to find kink- and hump-type solitary waves solutions 
as well as periodic solutions in terms of the Jacobi elliptic 
sine and cosine functions. 
Indeed, using our symbolic code 
\verb|PDESpecialSolutions.m| 
\cite{baldwin-etal-jsc-2004} 
for the $\tanh$, ${\mathrm{sech}}$, and~Jacobi elliptic function methods, 
at a click of a button, one obtains simple exact solutions, which we cover first.
\vskip 4pt
\noindent
{\bf Case I}: The focusing Gardner equation $(\beta > 0)$.
\vskip 2pt
\noindent 
The most well-known solutions are
\begin{equation}
\label{sechsolsfocgardner}
u(x,t) = 
 - \left (\tfrac{\alpha}{2\beta} \pm \sqrt{\tfrac{6}{\beta}} k \, 
{\mathrm{sech}}\, \theta \right),
\end{equation}
with $\theta = k x - (k^2 - \tfrac{\alpha^2}{4 \beta}) k t + \delta$, 
and
\begin{equation}
\label{cnsolfocgardner}
u(x,t) = 
- \left( \tfrac{\alpha}{2\beta} \pm \sqrt{\tfrac{6}{\beta}} k \sqrt{m} \, 
{\mathrm{cn}} (\theta ; m)  \right), 
\end{equation} 
with 
$\theta = k x + \left( (1-2m) k^2 + \tfrac{\alpha^2}{4 \beta} \right) k t + \delta$. 
Both solutions for the minus sign (in front of the square root)
are shown in Figure~\ref{graph-gardnerfocussechcnsols}. 
Observe that as the value of $m$ gets closer to $1$, the~${\mathrm{cn}}$-function 
starts taking the shape of a ${\mathrm{sech}}$-profile. 
\vskip 0.00001pt
\noindent
% FIGURE 2
\vskip 0.00001pt
\noindent
Of course, (\ref{sechsolsfocgardner}) also follows directly from 
(\ref{mkdvcoshsolU0zero}) upon application of the 
transformation (\ref{gardnertfmkdv}). 
Using (\ref{mkdvcoshsolU0nonzero}) in the same way yields
\begin{equation}
\label{gardnercoshsolU0nonzero}
u(x,t) =  
U_0 - \frac{\alpha}{2\beta} 
    + \frac{3 k^2}{\beta 
      \left( U_0 + \sqrt{U_0^2 + \frac{3 k^2}{2 \beta}} \, 
      \cosh \theta 
      \right)},
\end{equation}
with 
$\theta = k x - (k^2 + \beta U_0^2 -\tfrac{\alpha^2}{4 \beta} ) k t + \delta$
and where $U_0$ is arbitrary. 
Setting $U_0 = \tfrac{\alpha}{2\beta}$ yields
\begin{equation}
\label{coshsolgardner}
u(x,t) = 
\frac{6 k^2}{\alpha (1 + \sqrt{1 + \tfrac{6 \beta}{\alpha^2} k^2} \,\cosh \theta)}, 
\end{equation}
with $\theta = k x - k^3 t + \delta = k (x - V t) + \delta$, 
where $V = k^2$ denotes the speed of the wave and 
$k^{-1}$ is the effective width of the solitary wave.
This special solution is frequently used in the literature~\cite{alejo-etal-tams-2012,grimshaw-studapplmath-2015,grimshaw-etal-chaos-2002,pelinovskii-slyunyaev-jetplett-1998,slyunyaev-pelinovskii-jetp-1999,wazwaz-cnsns-2007} 
and could also be computed via a Darboux transformation 
(see, e.g.,~\cite{slyunyaev-pelinovskii-jetp-1999}).
% \cite{saha-etal-springer-2022,slyunyaev-pelinovskii-jetp-1999}). 
For $\beta > 0$, (\ref{coshsolgardner}) is valid for 
all values of $V$, and since $V > 0$ the wave is travelling to the right.
\vskip 4pt
\noindent
{\bf Case II}: The defocusing Gardner equation ($\beta < 0 $).
\vskip 2pt
\noindent
The simplest exact solutions are
\begin{equation}
\label{tanhsolsdefocgardner}
u(x,t) = 
- \left( 
\tfrac{\alpha}{2\beta} \pm \sqrt{-\tfrac{6}{\beta}} k \,
\tanh \theta \right),
\end{equation}
with $\theta = k x + (2 k^2 + \tfrac{\alpha^2}{4 \beta}) k t 
+ \delta$, 
and
\begin{equation}
\label{snsolsdefocgardner}
u(x,t) = 
 - \left( \tfrac{\alpha}{2\beta} \pm \sqrt{-\tfrac{6}{\beta}} k \sqrt{m} \, 
 {\mathrm{sn}} (\theta ; m) \right), 
\end{equation} 
with 
$\theta = k x + 
\left( (1+m) k^2 + \tfrac{\alpha^2}{4 \beta} \right) k t + \delta$.
In each of these solutions, $\delta$ is an arbitrary constant phase.
The shock wave solution (\ref{tanhsolsdefocgardner}) and periodic solution (\ref{snsolsdefocgardner}) can be found in, e.g.,~ref.~\cite{ono-jphyssocjpn-1976}.
Both solutions for the minus sign (in front of the square root)
are shown in Figure~\ref{graph-gardnerdefocustanhsnsols}. 
As the value of $m$ draws closer to $1$, the~${\mathrm{sn}}$-function starts taking 
the shape of a ${\mathrm{tanh}}$-profile. 
\vskip 0.00001pt
\noindent
% FIGURE 3
\vskip 0.00001pt
\noindent 
With respect to (\ref{coshsolgardner}) where $V = k^2$, 
note that when $\beta <0$, the argument of the square root, 
$1 + \tfrac{6 \beta}{\alpha^2} k^2$, 
will be zero when $V = V_{\mathrm{crit}} = \tfrac{\alpha^2}{6|\beta|}$.
Thus, (\ref{coshsolgardner}) is only valid when the speed is below that 
critical value ($ V < V_{\mathrm{crit}}$). 
Thus, when dealing with the defocusing Gardner equation, if~
$V \ge V_{\mathrm{crit}}$,
there is no solution of type (\ref{coshsolgardner}). 
Turning the argument around, solutions for (\ref{coshsolgardner}) 
of large amplitude (which is proportional to $\tfrac{k^2}{\alpha}$) or 
fast traveling waves (since $V = k^2$) can only occur if $|\beta|$ 
is relatively small in comparison with $\alpha$.
For example, for~$k = 2$, one must require that 
$|\beta| < \tfrac{\alpha^2}{24}$.

\begin{figure}[H]

\begin{tabular}{ccc}
\includegraphics[width=2.1875in,height=1.9625in]{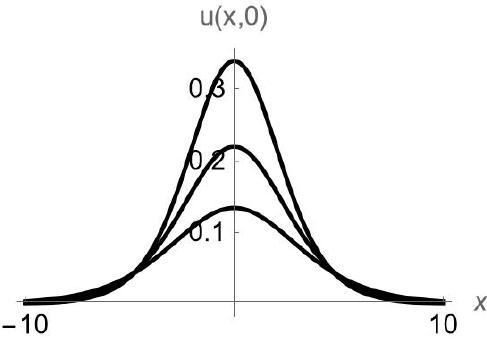}
& \phantom{x} & 
\includegraphics[width=2.1875in,height=1.9625in]{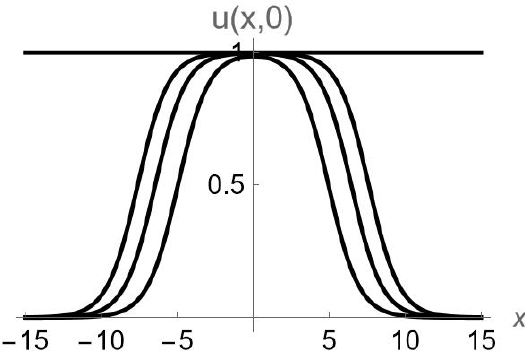}
\end{tabular}

% MDPI: Some content is overlap. please check if it affect reading. 
% if so, please revise.
% WH: Good the way it is. Made left and right bold as in the other figures. 
% WH: Added a line to explain which curve corresponds to which value of $k$.
\caption{Graphs of (\ref{coshsolgardnerspecial}) for 
$k = 0.5$, $k = 0.625$, and~$k = 0.75$ (\textbf{left}), 
and~$k=0.9999$, $k = 0.999995$, $k = 0.9999995$, 
and~$k = 1$ (\textbf{right}). The solitary wave becomes taller and narrower 
as the value of $k$ increases. }
\label{graph-gardnerSolutionsGrimshaw}
\end{figure}

\begin{figure}[H]

%\includegraphics[width=4.75in,height=2.75in]{2DGardnerDefocusCoshvsKinkAntikinkSols-new.pdf}

% WH: figure replaced. New figure has 0.9999 instead of .9999
\includegraphics[width=4.75in,height=2.75in]{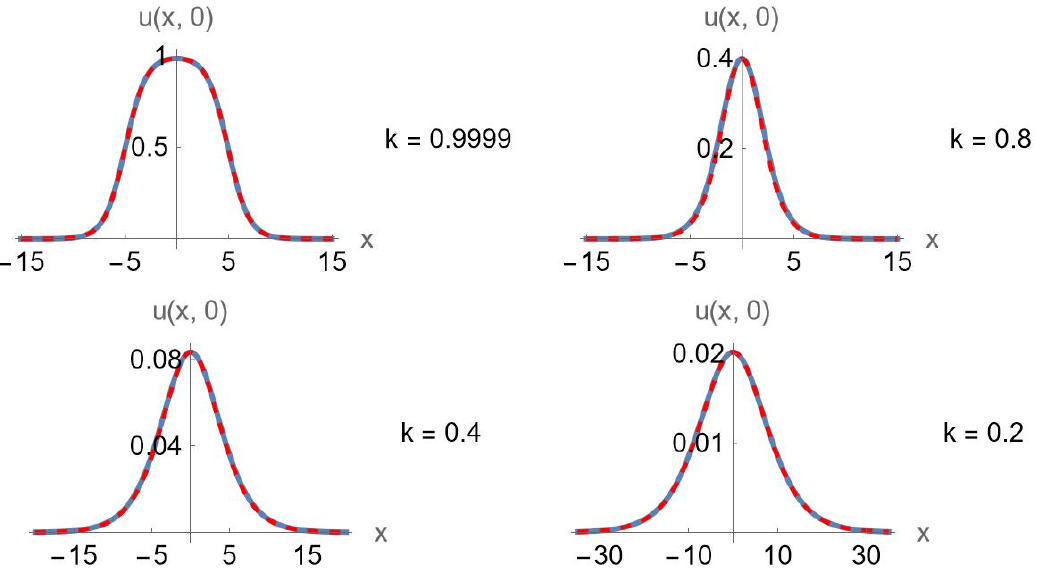}

% MDPI: Please complete 0 before digital dot, e.g.,~.9999 should be 0.9999.
% WH: New figure provided.  
\caption{Graphs of (\ref{coshsolgardnerspecial}) (full line) and (\ref{gardnerkinkantikink}) (dashed line) for four different values of $k$.}
\label{graph-comparisoncoshkinkantikink}
\end{figure}

The graphs in Figure~\ref{graph-gardnerSolutionsGrimshaw} are for 
$\alpha = 6$, $\beta = -6$, and~$\delta = 0$
(i.e., $ V < V_{\mathrm{crit}}$ is equivalent to $k^2 < 1$), 
for which (\ref{coshsolgardner}) simplifies into
\begin{equation}
\label{coshsolgardnerspecial}
u(x,t) = \frac{k^2}{1 + \sqrt{1-k^2}\,\cosh(k x - k^3 t)}. 
\end{equation}
As the value of $k$ increases, the solitary wave get taller and narrower. 
At $t = 0$ and values of $k$ very close to $1$, the waves become flat 
at the top, hence the~name table-top (or flat-top) waves. 
For the critical value $k=1$, the~wave degenerates into a horizontal line 
corresponding to
\begin{equation}
\label{gardnerGrimshawsollimit}
\lim_{k \to 1}\, \Bigl( \frac{k^2}{1 + \sqrt{1-k^2}\,\cosh(k x)} \Bigr) = 1.
\end{equation} 
For values of $k$ near $1$, solution (\ref{coshsolgardnerspecial}) 
can be very well approximated by a kink--antikink pair~\cite{pelinovskii-slyunyaev-jetplett-1998},
\begin{equation}
\label{gardnerkinkantikink}
u(x,t) = \tfrac{1}{2} 
 \left( \tanh \left( \tfrac{k}{2} (x - k^2 t + \Delta) \right) 
  - \tanh \left( \tfrac{k}{2} (x - k^2 t - \Delta) \right) \right),
\end{equation}
where $\Delta = 2 k^{-1} {\mathrm{Arctanh}}(1 - \sqrt{1 - k^2})$
serves as a measure for the width of the table-top wave. 
These kink and anti-kink solutions, which correspond to the 
left and right flanks of the table-top solutions, 
are clearly visible in Figure~\ref{graph-gardnerSolutionsGrimshaw} 
where $k$ approaches $1$. 
Although the expressions do not match analytically, 
Figure~\ref{graph-comparisoncoshkinkantikink} shows that the graphs of (\ref{coshsolgardnerspecial}) and (\ref{gardnerkinkantikink}) 
for $t=0$ nearly overlap even when $k$ is not close to $1$.  

Parenthetically, Grosse \cite{grosse-lettmathphys-1984} (Equation (24)) computed a two-soliton solution of the defocusing mKdV equation. 
His analytic solution supposedly describes the interaction of two kink-solutions.
It does not satisfy the equation exactly, but appears to be an excellent  
approximation to a solution, and therefore warrants further~investigation.

% FIGURE 4
\vskip 0.00001pt
\noindent
% WH: Oct 2, 2024 figure replaced. New figure has 0.9999 instead of .9999
% FIGURE 5

Notice that all the above solutions follow the scaling homogeneity 
(\ref{gardnerscale}).
Functions like ${\mathrm{cosh}}$, $\tanh$, ${\mathrm{cn}}$, etc.,
have no weights. 
With regard to the weights (\ref{gardnerweights}), 
$W(\tfrac{\alpha}{2\beta}) = 1$, as it should, because $W(u) = 1.$
Furthermore, 
$W(k) = 1$ because $W(x) = -1$ and $W(k x) = 0$. 
All terms in any $\theta$ must have weight zero, in~particular, $W(\delta) = 0$, and $W(m) = 0$, where $m$ is the modulus of any of the Jacobi elliptic functions.
From $W(t) = -3$, it follows that $W(V)= 2$ and $W(\omega) = 3$, 
where $\omega$ is the angular frequency in 
$\theta = k x - \omega t + \delta = k (x - V t) + \delta$.
Hence, if~$\omega$ and $V$ are polynomials in $k$, then $\omega$ can only have terms 
proportional to $k^3, \alpha k^2,$ and $\alpha^2 k$, and 
$V$ can only have terms in $k^2, \alpha k$, and~$\alpha^2$. 
The proportionality factors could have any powers of $\beta$ since $W(\beta) = 0$.

% SOLITONS
\section{Soliton~Solutions}
\label{gardner-solitons}
\begin{figure}[H]

\begin{tabular}{ccc}
\includegraphics[width=1.52in, height=1.76in]{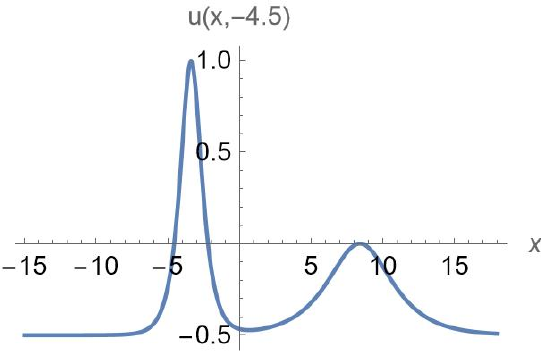} &
\includegraphics[width=1.52in, height=1.76in]{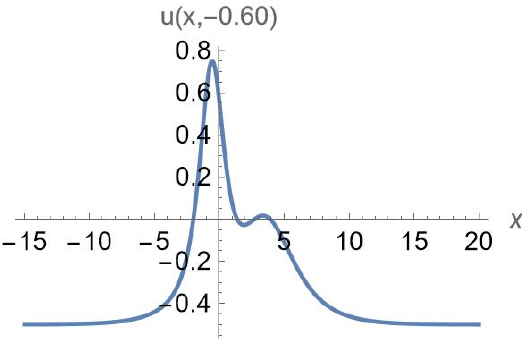} &
\includegraphics[width=1.52in, height=1.76in]{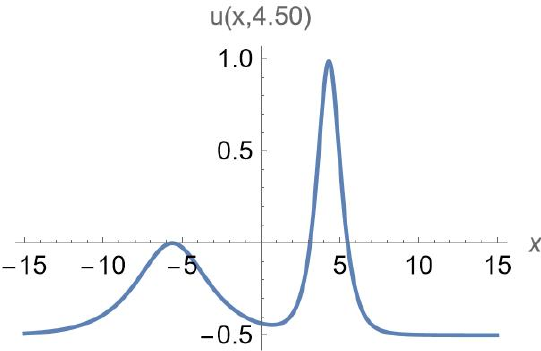}
\end{tabular}

% MDPI: Some content is overlap. please check if it affect reading. 
% if so, please revise.
% WH: all good as is. 
\caption{Graph of the two-soliton solution (\ref{gardnertwosoliton}) 
of the focusing Gardner equation at three different moments in time. }
\label{gardnerfocus-two-solitons-in-time}
\end{figure}

\begin{figure}[H]

%\includegraphics[width=2.90in,height=2.55in]{3DGardnerFocusCollisionTwoSolitons.pdf}

% WH: figure 20% smaller so that Case II start on same page 
\includegraphics[width=2.32in,height=2.04in]{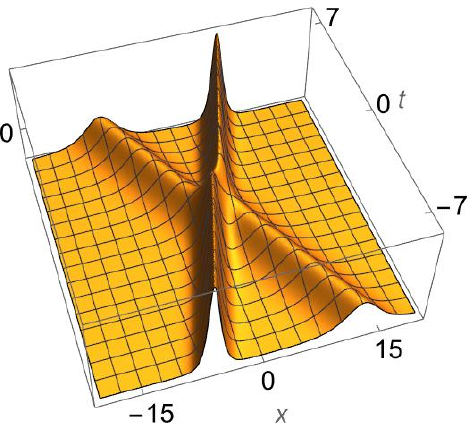}

\caption{Bird's eye view of a two-soliton collision for the focusing 
Gardner equation. Notice the phase shift after collision: the taller (faster) soliton is shifted forward and the shorter (slower) soliton backward relative to where they would have been if they had not collided. }
\label{graph-gardnerfocustwosolitoncollision}
\end{figure}

When considering solitons we again must make the distinction between the 
focusing and defocusing Gardner equations. 
\vskip 2pt
\noindent
{\bf Case I}: The focusing Gardner equation ($\beta > 0 $).
\vskip 2pt
\noindent
Using (\ref{gardnertfmkdv}), solutions $u(x,t)$ of the focusing 
version of (\ref{gardner}) are given by
\begin{equation}
\label{gardnersolugeneral}
u(x,t) = U(X(x,t), T(t)) - \tfrac{\alpha}{2\beta}, 
\end{equation}
where $U(X,T)$ is any soliton solution of the focusing mKdV equation and
$X(x,t) = x + \tfrac{\alpha^2}{4\beta}\, t $ and $T = t$. 
For example, the~two-soliton solution of (\ref{gardner}) reads
\begin{equation}
\label{gardnertwosoliton}
u(x,t) = - \frac{\alpha}{2\beta}
  + 2 \sqrt{\frac{6}{\beta}}
  \left(
  \frac{k_1 {\mathrm{e}}^{\theta_1} + k_2 {\mathrm{e}}^{\theta_2}
  + a_{12} \,
    (k_1 {\mathrm{e}}^{\theta_2} + k_2 {\mathrm{e}}^{\theta_1})
        {\mathrm{e}}^{\theta_1 + \theta_2}}{
    1 + {\mathrm{e}}^{2\theta_1} + {\mathrm{e}}^{2\theta_2}
  + \displaystyle{\frac{8 k_1 k_2}{(k_1 + k_2)^2}}\,  
  {\mathrm{e}}^{\theta_1 + \theta_2}
  + a_{12}^2 \, {\mathrm{e}}^{2\theta_1 + 2\theta_2}} 
  \right),
\end{equation}
where 
$\theta_i = k_i x - (k_i^2 - \tfrac{\alpha^2}{4\beta}) k_i t + \delta_i$ 
and 
$a_{12} = \left( \frac{k_1-k_2}{k_1+k_2} \right)^2$.
The elastic scattering of two solitons for the focusing Gardner equation 
is shown in Figures~\ref{gardnerfocus-two-solitons-in-time} and~\ref{graph-gardnerfocustwosolitoncollision},  
which have 2D and 3D graphs
of (\ref{gardnertwosoliton}) for $\alpha = \beta = 6$ 
with $k_1 = \tfrac{3}{2}$, $k_2 = \tfrac{1}{2}$, and~$\delta_1 = \delta_2 = 0$.
\vskip 0.001pt
\noindent
% FIGURE 6
\unskip

% WH: Oct 2, 2024 figure 20% smaller so that Case II start on same page 
% FIGURE 7

The Gardner equation has solitons of all orders $N$, which confirms
once more that it is a completely integrable~PDE.

\begin{figure}[H]

\begin{tabular}{ccc}
\includegraphics[width=2.1975in,height=1.9715in]{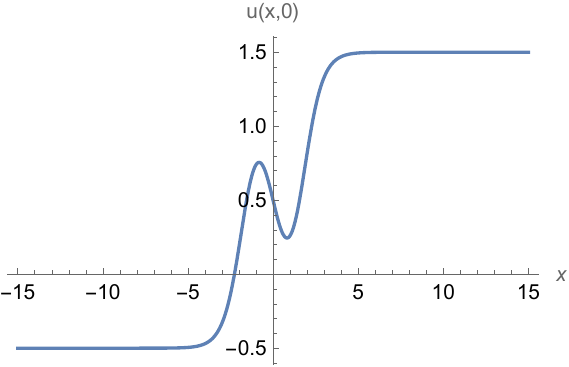}
& \phantom{x} & 
\includegraphics[width=2.1875in,height=1.9625in]{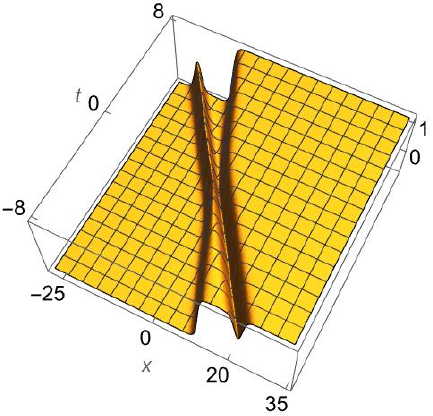}
\end{tabular}

\caption{2D and 3D graphs of solution (\ref{gardnerdefoctwosolitonMucalica})
for $\alpha = 6$, $\beta = -6$, $\delta = 0.65$ and $x_0 = 0$. }
\label{Graph-twosolitonGardnerMucalicadeltapt65}
\end{figure}

\begin{figure}[H]

\begin{tabular}{ccc}
\includegraphics[width=2.1975in,height=1.9715in]{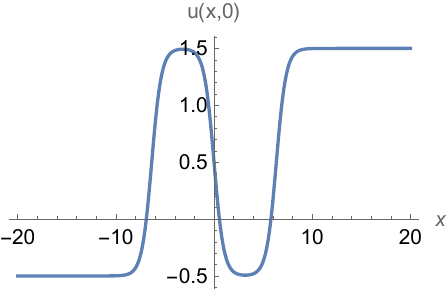}
& \phantom{x} & 
\includegraphics[width=2.1875in,height=1.9625in]{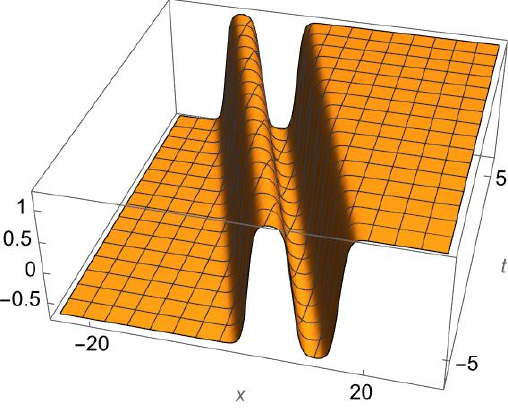}
\end{tabular}

\caption{2D and 3D graphs of solution (\ref{gardnerdefoctwosolitonMucalica})
for $\alpha = 6$, $\beta = -6$, $\delta = 3$ and $x_0 = 0$. }
\label{Graph-twosolitonGardnerMucalicadelta3}
\end{figure}

In general, the~existence of a two-soliton solution is no guarantee that the given PDE is completely integrable, but the existence of three-soliton solutions is a good indicator that the PDE is completely integrable.
Indeed, as~shown in~\cite{hereman-goktas-springer-2024}, there are PDEs that have at most two-soliton solutions but no three-soliton solutions. 
% 
% \vfil
% \newpage
\vskip 1pt
\noindent
{\bf Case II}: The defocusing Gardner equation ($\beta < 0 $).
\vskip 2pt
\noindent
Based on (\ref{mkdvdefoctwosol}), 
one obtains a two-soliton solution
\begingroup
\makeatletter\def\f@size{8}\check@mathfonts
\def\maketag@@@#1{\hbox{\m@th\normalsize\normalfont#1}}%
\begin{eqnarray}
\label{gardnerdefoctwosolitonMucalica}
\phantom{.}\!\!\!\!\!\!\!&&\!\! u(x,t) = - \frac{\alpha}{2 \beta} 
\nonumber \\
\!\!\!\!\!\!\!\!&&\!\!\!\! + \sqrt{\!-\frac{6}{\beta}}
 \left(\! 
 \frac{
 \sinh(\theta \!+\! 2\delta){\mathrm{e}}^{-2\eta}
 \!+\! \sinh(\theta \!-\! 2\delta){\mathrm{e}}^{2\eta}
 \!+\! 2 \sinh(\theta)(1 \!-\! \sinh^2(2\delta)){\mathrm{sech}}(2\delta)
 }{
 \cosh(\theta \!+\! 2\delta) {\mathrm{e}}^{-2\eta}
 \!+\! \cosh(\theta \!-\! 2\delta){\mathrm{e}}^{2\eta}
 \!+\! 2 \cosh(\theta)\cosh(2\delta) 
 }\!\right), 
\end{eqnarray}
\endgroup
with 
$\theta = x + \left( 2 + \frac{\alpha^2}{4 \beta} \right) t,$ 
and 
$\eta = 
\left( x + \left( 
       2 + \frac{\alpha^2}{4 \beta} + 4 {\mathrm{sech}}^2 (2\delta) 
       \right) t + x_0 
 \right) \tanh(2\delta),$
with $\delta > 0$ and $x_0$ arbitrary real parameters.
Solution (\ref{gardnerdefoctwosolitonMucalica}), which describes the coalescence of 
two wave fronts, is pictured in 2D and 3D in 
Figures~\ref{Graph-twosolitonGardnerMucalicadeltapt65} and~\ref{Graph-twosolitonGardnerMucalicadelta3} for two different values of $\delta$.
\vskip 0.00001pt
\noindent
% FIGURE 8
\vskip 0.00001pt
\noindent
% FIGURE 9
\unskip

%%%%%%%%%%%%%%%%%%%%%%%%%%%%%%%%%%%%%%%%%%
% SOFTWARE   
\section{Symbolic~Software}
\label{symbolic-software}
Using the concept of scaling homogeneity, we have been able 
to create powerful algorithms to investigate the complete integrability of systems of polynomial nonlinear PDEs. 
In this section, we give a brief summary of the available~codes. 

Our {\em Mathematica} code \verb|PainleveTest.m| \cite{baldwin-hereman-code-painlevetest-2001} 
automates the Painlev\'{e} test, which allows one to verify if a nonlinear PDE 
has the Painlev\'e property~\cite{baldwin-hereman-jnlmp-2006} as discussed in Section~\ref{painleve-analysis}.   

The {\em Mathematica} code \verb|InvariantsSymmetries.m| \cite{goktas-hereman-code-invarsym-1997}
% \cite{goktas-hereman-pd-1998} 
computes polynomial conserved densities and higher-order symmetries of 
nonlinear $(1+1)$-dimensional PDEs that can be written as a polynomial 
system of evolution equations. 
If a PDE has arbitrary parameters, the code allows one to derive conditions on 
these parameters so that the PDE admits conserved quantities or 
generalized symmetries.
% To code also covers nonlinear differential-difference equations 
An example of such a ``classification'' problem is given in~\cite{goktas-hereman-jsc-1997}. 
A discussion of the scope and limitations of the code can be found in~\cite{goktas-hereman-jsc-1997,goktas-hereman-acm-1999}. 

\textls[-15]{To cover conservation laws of nonlinear PDEs in more than one space }\mbox{variable~\cite{hereman-ijqc-2006,poole-hereman-aa-2010,poole-hereman-jsc-2011}}, 
we developed 
\verb|ConservationLawsMD.m| \cite{poole-hereman-code-conslawsmd-2009}, 
a {\em Mathematica} package to compute polynomial conservation laws of polynomial 
systems of nonlinear PDEs in space variables $(x,y,z)$ and time $t$.

In~\cite{baldwin-hereman-ijcm-2010}, the~authors show details of the 
algorithm to compute recursion operators for systems of 
nonlinear PDEs of type (\ref{scalarpde}), including formulas for handling integro-differential 
operators used in Section~\ref{recursion-operator}. 
The {\em Mathematica} package \verb|PDERecursionOperator.m| 
\cite{baldwin-hereman-code-recoper-2003} performs the symbolic computation 
of recursion operators of systems of polynomial nonlinear evolution~equations. 
 
In Appendix B of~\cite{larue-msthesis-2011}, Larue presents 
\verb|LaxpairTester.m|, a~{\em Mathematica} code to verify Lax pairs 
in operator and matrix form. 
In~\cite{hickman-etal-aa-2012}, an~algorithm is presented to compute 
Lax pairs in operator form, but that algorithm has not been implemented~yet. 

In addition, we developed the {\em Mathematica} package
\verb|PDESpecialSolutions.m| 
\cite{baldwin-hereman-code-exactsols-2003}
to compute solitary wave solutions based on the $\tanh$-method and generalizations 
for the ${\mathrm{sech}}$, ${\mathrm{sn}}$, and~${\mathrm{cn}}$ functions
~\cite{baldwin-etal-jsc-2004}.

Recently, we added the {\em Mathematica} code \verb|PDESolitonSolutions.m| 
\cite{goktas-hereman-code-solitons-2023} 
to compute soliton solutions of polynomial PDEs based on a simplified version 
of Hirota's method described in~\cite{hereman-goktas-springer-2024}.

Since our codes only use tools from calculus, linear algebra, the~calculus 
of variations, and~differential geometry, these algorithms are fairly straightforward 
to implement in the syntax of computer algebra systems such as {\em Mathematica}, 
{\em Maple}, and~{\em REDUCE}.
Our software is open source and available in the public domain. 
All our {\em Mathematica} packages and notebooks are available 
on the Internet at 
% MDPI: Please add the access date (Format: Date Month Year). 
% e.g.,~(accessed on 1 January 2020).
% WH: Done. 
\url{https://people.mines.edu/whereman/} 
(accessed on 1 October 2024). 
% WH 09/22/2024 reference to Wolfram site added
A summary of the codes used in this paper can also be 
found at 
% MDPI: Please add the access date (Format: Date Month Year). 
% e.g.,~(accessed on 1 January 2020).
% WH: done 
\url{https://community.wolfram.com/groups/-/m/t/3275116} 
(accessed on 23 September 2024).

%%%%%%%%%%%%%%%%%%%%%%%%%%%%%%%%%%%%%%%%%%
% CONCLUSIONS
\section{Conclusions and Future~Work}
\label{conclusions-future-work}
The approach described in this paper and related software is applicable to 
large classes of nonlinear PDEs, which can be expressed as polynomial systems of 
evolution equations. 
As a prototypical example, we gave a detailed integrability analysis of the Gardner equation by computing its densities (and fluxes), higher-order symmetries, 
recursion operator, Lax pair, Hirota's bilinear representation, and~soliton solutions.
The corresponding results for the KdV and mKdV equations were obtained by setting 
the coefficient of the cubic and quadratic term equal to zero, respectively. 

We also showed how to compute the Gardner (Miura, resp.) transformation, which 
connects solutions of the Gardner (mKdV, resp.) equation to those of the KdV equation. 
With the Gardner transformation, some solutions of the KdV equation could be 
obtained from those of the Gardner equation shown in this paper. 
Likewise, applying the Miura transformation to solutions of the mKdV equation will 
lead to solutions of the KdV equation.  
When new solutions of the Gardner and mKdV equations are discovered, 
it would be worthwhile to investigate which solutions of the KdV equation they 
correspond to and, more importantly, if~new solutions of the KdV equation 
could be computed that way 
(see, e.g.,~ref.~\cite{kirane-etal-csf-2024}).

The crux of our computational strategy is a skillful use of the 
scaling symmetry of the PDE and relies on the observation 
that the defining equations for conservation laws, 
generalized symmetries, recursion operator, Lax pair, bilinear 
representation, and~Gardner transformation should only hold on solutions of the given PDE.
Consequently, the~quantities (or operators) one computes inherit the 
scaling symmetry of the given~PDE. 
 
Since their defining equations are similar, it would also be possible to use this 
approach to compute symplectic and Hamiltonian (co-symplectic) operators of PDEs. 
In doing so, it would be possible to verify whether or not a PDE has a bi-Hamiltonian 
(or tri-Hamiltonian) structure, which is yet another criterion for its complete integrability. 
Further exploration of this idea, as well as the design of algorithms and codes 
for the computation of symplectic and Hamiltonian operators, is left for future~work. 

The methodology discussed in this paper might also apply to the Gardner equation 
in  $(2+1)$ dimensions~\cite{konopelchenko-book-1992,zhou-ma-nuovocimentob-2000}, 
which is a combination of the Kadomtsev--Petviashvili (KP) and the modified KP~equations.

The algorithms used in this paper are coded in {\em Mathematica} syntax,
but can be adapted for major computer algebra systems such as 
{\em Maple} and {\em REDUCE}. 

\authorcontributions{
% MDPI: For research articles with several authors, 
% a~short paragraph specifying their individual contributions must be provided. 
% The~following statements should be used ``Conceptualization, X.X. and Y.Y.; 
% methodology, X.X.; software, X.X.; validation, X.X., Y.Y. and Z.Z.; 
% formal analysis, X.X.; investigation, X.X.; resources, X.X.; data curation, 
% X.X.; writing---original draft preparation, X.X.; writing---review 
% and editing, X.X.; visualization, X.X.; supervision, X.X.; project 
% administration, X.X.; funding acquisition, Y.Y. 
% All authors have read and agreed to the published version of the 
% manuscript.'', please turn to the 
% \href{http://img.mdpi.org/data/contributor-role-instruction.pdf}{CRediT 
% taxonomy} for the term explanation. 
% Authorship must be limited to those who have contributed substantially 
% to the work~reported.
% WH: done. 
% 
Both authors contributed equally to this work. 
Conceptualization, W.H.; methodology, W.H.; software, U.G., validation, U.G., 
formal analysis, W.H. and U.G.; investigation, W.H. and U.G.; 
resources, W.H. and U.G.; writing -- original draft preparation, W.H.; 
writing -- review and editing, W.H. and U.G.; visualisation, U.G.; 
supervision, W.H.; project administration, W.H.; funding acquisition W.H. 
All authors have read and agreed to the published version of the manuscript}
\funding{The material presented in this paper is based in part upon research 
supported by the National Science Foundation (NSF) of the United States of America 
under Grants Nos. CCR-9300978, CCR-9625421, CCR-9901929, and~CCF-0830783.
Any opinions, findings, and~conclusions or recommendations expressed in this material 
are those of the authors and do not necessarily reflect the views of~NSF.
}
\dataavailability{Our software is open source and in the public domain. 
All our {\em Mathematica} packages and notebooks are available on the Internet 
at
% MDPI: Please add the access date (Format: Date Month Year). 
% e.g.,~(accessed on 1 January 2020).
\url{https://people.mines.edu/whereman/} (accessed on 2 September 2024).
}

\acknowledgments{
The authors are grateful to Frank Verheest 
(Sterrenkundig Observatorium, Univ.\ Gent, Belgium) for valuable 
discussions on the use of the Gardner equation in plasma physics.
We thank him, Rehana Naz (Dept.\ Math.\ Stat.\ Sci., 
Lahore School of Economics, Pakistan), 
Wen-Xiu Ma (Dept.\ Math.\ Stat.,
Univ.\ South Florida, Tampa, FL),  
and the anonymous referees for their thoughtful comments, 
suggestions, and~references which helped further improve 
the~paper. 
}
\conflictsofinterest{The authors declare no conflicts of~interest.}
%%%%%%%%%%%%%%%%%%%%%%%%%%%%%%%%%%%%%%%%%%
% REFERENCES BIBLIOGRAPHY 

\begin{adjustwidth}{-\extralength}{0cm}
% \printendnotes[custom] % Un-comment to print a list of endnotes

\reftitle{References}

% Please provide either the correct journal abbreviation 
% (e.g. according to the “List of Title Word Abbreviations” 
% http://www.issn.org/services/online-services/access-to-the-ltwa/) 
% or the full name of the journal.
% Citations and References in Supplementary files 
% are permitted provided 
% that they also appear in the reference list here. 

% \bibliography{your_external_BibTeX_file}

%=====================================
% References, variant B: internal bibliography
%=====================================
  
% TO DO adjust all reference to correct format 
% BIBLIO 

\PublishersNote{}
\end{adjustwidth}
\end{document}